\documentclass[12pt,preprint]{aastex}
\usepackage{amsmath}
\usepackage{longtable}
\usepackage{subfigure}

\shorttitle{Turbulent Cascade to Explain GRB Spectra}

\shortauthors{Mao, Li, \& Wang}

\begin{document}


\title{Spectral Diversities of Gamma-ray Bursts in High Energy Bands: Hints from Turbulent Cascade}


\author{
Jirong Mao\altaffilmark{1,2,3}, Liande Li\altaffilmark{1,2,4}, and Jiancheng Wang\altaffilmark{1,2,3}
}
\altaffiltext{1}{Yunnan Observatories, Chinese Academy of Sciences, 650011 Kunming, Yunnan Province, China}
\altaffiltext{2}{Center for Astronomical Mega-Science, Chinese Academy of Sciences, 20A Datun Road, Chaoyang District, Beijing, 100012, China}
\altaffiltext{3}{Key Laboratory for the Structure and Evolution of Celestial Objects, Chinese Academy of Sciences, 650011 Kunming, China}
\altaffiltext{4}{University of Chinese Academy of Sciences, 100049 Beijing, China}

\email{jirongmao@mail.ynao.ac.cn}

\begin{abstract}
We statistically examine the gamma-ray burst (GRB) photon indices obtained by the Fermi-GBM and Fermi-LAT observations and compare the LAT GRB photon indices to the GBM GRB photon indices. We apply the jitter radiation to explain the GRB spectral diversities in the high-energy bands. In our model, the jitter radiative spectral index is determined by the spectral index of the turbulence.
We classify GRBs into three classes depending on the shape of the GRB high-energy spectrum when we compare the GBM and LAT detections:
the GRB spectrum is concave (GRBs turn out to be softer and are labeled as S-GRBs), the GRB spectrum is convex (GRBs turn out to be harder and are labeled as H-GRBs), and the GRBs have no strong spectral changes (labeled as N-GRBs). A universal Kolmogorov index $7/3$ in the turbulent cascade is consistent with the photon index of the N-GRBs. The S-GRB spectra can be explained by the turbulent cascade due to the kinetic magnetic reconnection with the spectral index range of the turbulence from $8/3$ to 3.0. The H-GRB spectra originate from the inverse turbulent cascade with the spectral index range of the turbulence from 2.0 to 3.5 that occurred during the large lengthscale magnetic reconnection. Thus, the GRB radiative spectra are diversified because the turbulent cascade modifies the turbulent energy spectrum. More observational samples are expected in the future to further identify our suggestions.
\end{abstract}


\keywords{Gamma-ray bursts --- Magnetic fields --- Magnetohydrodynamics}


\section{Introduction}
Gamma-ray bursts (GRBs) have been detected by some astronomical satellites, such as Fermi and Swift. Fermi-GBM and Fermi-LAT have the detection energy ranges of 8 keV$-$40 MeV and 30 MeV$-$300 GeV, respectively, while Swift-BAT has the detection energy range between 15 and 350 keV. The third GBM catalog \citep{bhat16}, the second LAT catalog \citep{ajello19}, and the third BAT catalog \citep{lien16} provided the GRB spectral information in detail. Usually, the GRB prompt spectrum can be fitted by the Band function, and three important parameters, which are low-energy photon index, high-energy photon index, and peak energy $E_{\rm{peak}}$, are included \citep{band93}.
It is interesting to investigate the GRB photon indices in the high-energy bands.

Synchrotron radiation can be one possible mechanism to reproduce GRB prompt emission \citep{piran04}.
When relativistic electrons accelerated by GRB shocks have a power-law energy distribution, the synchrotron radiation spectrum has a power-law shape, and the photon index is determined by the index of the electron energy distribution.
This scenario is adopted to explain the observed GRB prompt spectra, although it does not fit very well to the low-energy photon indices of some GRBs (Preece et al. 1998, see also further discussion in Section 4). Moreover, the LAT GRBs have a time delay to the GBM GRBs, and the LAT-detected emission has usually been proposed to be original from the GRB afterglow with the radiation mechanism of inverse Compton (IC) or synchrotron self-Compton (SSC; Ajello et al. 2019).
Besides the successful issues mentioned above, some challenges are shown. For example, if GRB outflow is magnetic-dominated, electrons will suffer an efficient cooling in a very rapid way by synchrotron radiation. \citet{be14} suggested that the magnetic-dominated outflows may have a process of energy dissipation before the synchrotron radiation. Proton synchrotron radiation is suggested to be one cooling mechanism in GRB prompt emission \citep{gh20}.

Each GRB temporal pulse includes many spikes that have very short timescales. GRB prompt emission can be constrained to occur in a small region under the very short timescale. Thus, the very short timescale indicates that the emission region is located at a very small lengthscale. The timescales were identified by the time-dependent probability density functions \citep{bhatt12}. The spike timescale was measured to be within a range of $10^{-6}$$-$$10^{-3}$ s in general. \citet{gol14} took the Swift-BAT sample and performed the wavelet analysis on the lightcurves, and the shortest timescale was measured to be 10 ms. The GRB lightcurves observed by Fermi were also examined, and the shortest timescale corresponds to the GRB emission region of $10^{13}$$-$$10^{14}$ cm in the case in which the bulk Lorentz factor is larger than 400 \citep{gol15}. The observed short timescale provides some constrains on the theoretical investigations.
For instance, the radiation features in the very short timescales were presented by numerical simulations \citep{yuan16}.
Both dynamical and radiative processes may happen in such small timescales and lengthscales. In particular, stochastic processes can be involved. It seems reasonable to consider jitter radiation, the radiation of the relativistic electrons in the random and small-scale magnetic field, to be one possible radiation mechanism for GRBs \citep{me99}. In the high-energy band, jitter radiation has a power-law spectral shape \citep{me00,kelner13}. We have a series of analytic works on the perturbation theory of the jitter radiation \citep{mao07,mao11,mao12,mao13,mao17}.
Both short acceleration timescale and short cooling timescale were estimated \citep{mao11}. In particular, the spectral index of the jitter radiation is determined by the spectral index of the turbulence if the turbulent energy spectrum has a power-law shape, and the photons with higher energy are emitted at smaller lengthscale \citep{mao11}. Thus, the jitter radiative photon index is directly linked to the spectral index of the turbulence. In the jitter radiation framework, we can utilize the turbulent cascade properties to study the GRB spectral properties in the high-energy bands.

Turbulence can be introduced in the hydrodynamic process, and the GRB radiation is associated to the turbulence \citep{kumar09,lazar09,narayan09,lemoine13}. Nonlinear energy cascade usually occurs in turbulent flows across a large lengthscale range. Direct energy cascade from large to small lengthscales shows a typical Kolmogorov spectrum that has a power-law shape, and the cascade is terminated at a certain small lengthscale as shown by an exponential cutoff in a turbulent energy spectrum. Moreover, inverse cascade from small to large lengthscales also happens in turbulent flows.
One of the recent developments in the high-energy astrophysics is to link microphysics to radiation.
Relativistic turbulence has been introduced in the synchrotron and SSC processes \citep{uzdensky18}. The turbulent energy cascade extended from the hydrodynamic/magnetohydrodynamics (MHD) lengthscale to the kinetic lengthscale has been investigated \citep{zhdankin17,comisso18}. In particular, the turbulence and the magnetic reconnection are physically connected at kinetic lengthscales \citep{franci17,lo17}. This research has been applied to study high-energy emissions from blazars and pulsar wind nebulae \citep{pe16,zhdankin19}. Moreover, the inverse cascade of the magnetic turbulence was identified in the nonhelical case \citep{bran15}. This inverse cascade was also shown in the relativistic MHD case \citep{zrake14}. \citet{franci17} mentioned the possible inverse cascade in the plasma turbulence induced by the magnetic reconnection.
A magnetic island merger can be one mechanism for the inverse cascade \citep{zhou19}.
In fact, \citet{na15} and \citet{pe18} performed some kinetic simulations and represented the merger in the case of the plasmoid-dominated relativistic magnetic reconnection. The global current sheet can crash into many plasmoids due to the tearing mode instability and the turbulence \citep{mallet17}. Therefore, we may consider the link between the turbulent cascade and the GRB radiative spectrum in the high-energy bands.

We present the GRB photon indices obtained by the LAT, GBM, and BAT detections in Section 2.1. In section 2.2, we illustrate the jitter radiative spectrum related to the turbulent energy spectrum, when we consider the turbulent cascade. The GRB photon indices obtained by LAT and GBM related to the turbulent spectra are explained in the jitter radiation framework.
We present our results in Section 3. Some issues, such as data fitting, electron pitch-angle, particle acceleration, turbulent eddy turnover time, and a few special cases on the turbulent cascade are discussed in Section 4. The conclusion is given in Section 5.

\section{Link between Spectral Indices in High-energy Bands and Turbulent Cascade}

\subsection{GRB Spectral Diversity in High-energy Bands}
GRB emission in the high-energy bands has been well detected by Swift-BAT, Fermi-GBM, and Fermi-LAT. GBM spectra are well fitted by the Band function \citep{band93}. LAT spectra can be fitted by a power-law model. In order to examine the spectral indices of the GRB emission in the high-energy bands, we utilize the GRB catalogs where the GRB spectral indices are given. The second Fermi-LAT GRB catalog has the latest results on the GRB observational properties, in which the photon index of each GRB is included \citep{ajello19}. Moreover, the GBM photon index is given in the third GBM GRB catalog \citep{bhat16}. We use the high-energy photon index determined by the Band function to be the GBM photon index. Thus, we can compare the photon index determined by GBM and that by LAT for each GRB. We list 138 GRBs in Table 1, and the photon indices of both LAT and GBM are given. A few GBM GRBs do not have enough photons above the peak energy to perform a reliable spectral fitting, and the high-energy photon indices have very large error bars when the spectra are fitted by the Band function. A few GBM GRBs have the spectra that have significant thermal components or strong time evolution, and these spectra fitted by the Band function also have the high-energy photon indices with large error bars. Thus, these high-energy photon indices are excluded in Table 1.

\citet{ajello19} provided two kinds of photon indices from Fermi-LAT. One kind of photon index is obtained in the LAT observing time interval. The other kind of photon index is obtained in the GBM observing time interval. We list two kinds of LAT photon indices in Table 1. If both the LAT photon index and the GBM photon index are obtained in the GBM observing time interval, they are simultaneous. Because GBM observing time interval and LAT observing time interval are different for a certain GRB, the LAT photon index obtained in the LAT observing time interval is different from that obtained in the GBM observing time interval for a certain GRB. We can compare the LAT photon index obtained in the LAT observing time interval to that obtained in the GBM observing time interval. The distributions are shown in the upper panel of Figure 1. It is clearly seen that two distributions with two peaks are similar. We then draw the plot of the LAT photon indices obtained in the GBM observing time interval versus the LAT photon indices obtained in the LAT observing time interval in the lower panel of Figure 1. It is shown that the two kinds of photon index are not well consistent.

In order to further examine the photon indices of LAT and GBM for each GRB, we classify three GRB cases with the energy from the GBM band toward the LAT band:
the GRB spectrum is concave (GRBs turn out to be softer and labeled as S-GRBs), the GRB spectrum is convex (GRBs turn out to be harder and labeled as H-GRBs), and the GRBs have no strong spectral changes (labeled as N-GRBs).
Because two kinds of LAT photon indices are listed in Table 1, we note two classifications labeled as 1 and 2 respectively. When we consider the LAT photon indices obtained in the LAT observing time interval, we have 25 S-GRBs, 41 H-GRBs, and 16 N-GRBs. When we consider the LAT photon indices obtained in the GBM observing time interval, we have 23 S-GRBs, 15 H-GRBs, and 9 N-GRBs.

We perform the statistical analysis. We first consider the photon index distributions of LAT (LAT observing time interval) and GBM. The total photon index distributions are shown in the upper left panel of Figure 2. Two distributions have large overlaps. We perform a gaussian fitting to the two distributions. The photon index distribution of LAT GRBs has the peak around $-2.11\pm 0.40$ and that of GBM GRBs has the peak around $-2.33\pm 0.38$. It seems from the statistical distributions that the photon index distributions of LAT and GBM are roughly consistent when we include the relatively large distribution span. We also perform the photon index distributions for N-GRBs, S-GRBs, and H-GRBs in the upper right, lower left, and lower right panels of Figure 2, respectively. We clearly see that the photon index distributions of the LAT GRBs and the GBM GRBs are separated.

We then consider the photon index distributions of LAT (GBM observing time interval) and GBM. The total photon index distributions are shown in the upper left panel of Figure 3. Two distributions have large overlaps. We perform a gaussian fitting to the two distributions. The photon index distribution of LAT GRBs has the peak around $-2.40\pm 0.52$ and that of GBM GRBs has the peak around $-2.33\pm 0.38$. It seems from the statistical distributions that the photon index distributions of LAT and GBM are roughly consistent when we include the relatively large distribution span. We also perform the photon index distributions for N-GRBs, S-GRBs, and H-GRBs in the upper right, lower left, and lower right panels of Figure 3, respectively. We clearly see that the photon index distributions of the LAT GRBs and the GBM GRBs are separated.

We show the GBM GRB photon indices versus the LAT GRB photon indices (either obtained in the LAT observing time interval or obtained in the GBM observing time interval) in Figure 4. It is clearly seen that the GBM GRB photon indices and the LAT GRB photon indices have no correlation in total. The photon indices of three GRB classes are plotted by different colors. The photon index distributions of N-GRBs and S-GRBs are distinguished in two regions in the plots.

Some GRBs have extended emissions detected in the LAT energy band. In order to identify the photon indices related to the GRB duration, \citet{ack13}
presented one plot that includes the LAT photon index obtained in the interval between the end of GBM to the end of LAT versus the GBM photon index, and the observational data were given from the first Fermi-LAT GRB catalog. The two photon index classes are not correlated, and the spectrum described by the photon index obtained in the interval between the end of GBM to the end of LAT is harder than that described by the GBM photon index. In this paper, we directly compare the LAT photon index to the GBM photon index for N-GRBs, S-GRBs, and H-GRBs, respectively, in Figure 4. Thus, the GRB spectral diversities are clearly shown.

We finally perform the statistics of the low-energy photon indices and the $E_{\rm{peak}}$ numbers obtained by the GBM for N-GRBs, S-GRBs, and C-GRBs, respectively. The results are shown in Figures 5$-$8.
The $E_{\rm{peak}}$ distribution is mainly below about $2\times 10^3$ keV, and a few GRBs have the $E_{\rm{peak}}$ values larger than $4\times 10^3$ keV. However, all S-GRBs have the $E_{\rm{peak}}$ values smaller than $10^3$ keV, and some N-GRBs have the $E_{\rm{peak}}$ values larger than $10^3$ keV. It seems that S-GRBs have smaller $E_{\rm{peak}}$ values and H-GRBs have larger $E_{\rm{peak}}$ values. More data samples in the future are necessary to further confirm this trend.

The Swift-BAT detection has a relatively narrow energy range, and each GRB spectrum is fitted by both power-law model and cutoff power-law model.
In Table 1, we list the power-law/cutoff power-law index obtained in the third Swift-BAT GRB catalog as reference values.
$E_{\rm{peak}}$ numbers of most BAT-detected GRBs are in the BAT energy range, and the GRB energy release seems to be detected by BAT \citep{lien16}.
Therefore, the power-law or the cutoff power-law fitting to the BAT GRBs might be affected by the spectral feature that is below $E_{\rm{peak}}$.
We do not aim to compare the BAT photon index to the LAT/GBM photon index in this paper.

\subsection{jitter Radiation and Turbulent Cascade}
The jitter radiation to study the GRB emission has been presented in detail by \citet{mao11}. Here, we adopt the jitter radiation to explain the GRB spectral properties in the high-energy bands.
In particular, the radiation spectrum has a power-law shape, and the radiative index is determined by the spectral index of the turbulence.
In this paper, the radiative photon index number is equal to the spectral index number of the turbulence.
Furthermore, the radiation frequency $\omega$ is related to the turbulent wavenumber $q$ as $\omega=\gamma^2cq$, where $\gamma$ is the electron Lorentz factor, and $c$ is the light speed. This means that higher energy radiation occurs at smaller lengthscale.
If the turbulence has different turbulent spectral indices during the turbulent cascade, the GRB photon indices are different in the different energy bands.

It is well known that the hydrodynamic/MHD turbulence provides the Kolmogorov spectral index $p/3$, where $p$ is the cascade number. \citet{she94} provided a hydrodynamic turbulent cascade as $p/9+2[1-(2/3)^{p/3}]$, and this is the modification of the simple Kolmogorov energy cascade. We adopted the number of $7/3$ in the former work \citep{mao11}. Here, we clearly see that
a universal turbulent energy spectrum cannot simply explain the GRB spectral diversities. If we propose that the turbulent cascade or inverse cascade can be one possible mechanism to explain the GRB spectra in the high-energy bands, we consider three cases below.
(1) The turbulent cascade process (either cascade or inverse cascade) simply keeps the turbulent spectral shape in one GRB, and the GRB has roughly the same photon index in both GBM and LAT energy bands. Thus, the GRB can be classified as N-GRBs.
(2) The turbulent cascade from large lengthscale to small lengthscale in one GRB is not a simple extension. The hydrodynamic cascade provided by \citet{she94} is dependent on the parameter $p$. Moreover, the hydrodynamic/MHD turbulence can be developed into the kinetic lengthscale \citep{sch09}, and the kinetic turbulent energy spectrum is deeper.
The spectral index of the kinetic turbulence sometimes has a narrow range between $7/3$ and 3 \citep{zhao16}. In particular, magnetic reconnection can drive plasma turbulence cascade and provide the spectral index range of the turbulence between $8/3$ and 3, with a typical value of 2.8 at the sub-ion lengthscale \citep{lo17}. Thus, we consider the turbulent cascade from large lengthscale to small lengthscale in one GRB, and the spectral index varies from a typical Kolmogorov number $5/3$ to be less than 3. One GRB turns out to be softer from the GBM energy band to the LAT energy band, and this GRB is classified as S-GRB.
(3) The turbulent inverse cascade from small lengthscale to large lengthscale was identified by some numerical simulations \citep{zrake14,bran15}. The inverse cascade may cause one turbulent energy spectrum to be flatter. For example, besides the cascade by the local interactions, the turbulent energy can be injected directly due to the magnetic reconnection. The reconnection can produce small magnetic islands, then these islands merge at large lengthscales. This process has been simulated by \citet{franci17}, and it also occurs in the plasmoid-dominated magnetic reconnection \citep{na15,pe18}.
\citet{zhou19} performed a further confirmation on this issue.
The hard turbulent spectral shapes due to the inverse cascade have been comprehensively presented by \citet{zrak16}.
Thus, one GRB with the inverse cascade turns to be harder from the GBM energy band to the LAT energy band, and this GRB is classified as H-GRB.

We note that two GRBs (GRB 100620A and GRB 100724B) detected by LAT have photon index numbers of about 4, although the error bars are relatively large. We notice the following case. The kinetic turbulent cascade of the relativistic plasma can reach the sub-Larmor lengthscale, and the spectral index of the turbulence can be a number of 4 \citep{zhdankin17,comisso18}. This indicates that some very soft GRBs are emitted due to the kinetic turbulent cascade at very small lengthscales. In addition, we note that GRB 100414A has a GBM photon index number of 3.53, and the error bar is large.

\section{Results}
The GRB photon index is determined by the spectral index of the turbulence in the jitter radiation framework. The LAT photon index and the GBM photon index for a certain N-GRB has no significant change. This indicates that a universal spectral index of turbulence is within the turbulent energy cascade. Compared to the statistics of the N-GRB photon index, the Kolmogorov index $7/3$ is consistent to the photon index of N-GRBs.

We propose that the turbulent cascade causes the GRB photon index variation in the high-energy bands. The S-GRBs turn out to be soft toward high-energy radiation frequency due to the turbulent cascade from large to small lengthscales. From the photon index statistics of the S-GRBs, most photon index numbers are in the range from 2.5 to 3.0. The turbulence cascade due to the magnetic reconnection provides the spectral index of the turbulence range from $8/3$ to 3.0 with a typical value of 2.8 \citep{lo17}. We think that this turbulent cascade mode can be one of the mechanisms to make the GRB spectrum soft toward high-energy radiation frequency.

We suggest that the inverse turbulent cascade occurs in the H-GRBs. The inverse cascade from small to large lengthscales provides a typical index of 2.0 in the turbulent energy spectrum \citep{zrake14,bran15}. The simulations in the condition of the large-scale magnetic reconnection provided a set of spectral index of turbulence with the range from 2.0 to 3.5 \citep{zhou19}. This set of inverse cascade index is consistent to the photon index statistics of the H-GRBs. Thus, the hard GRB spectrum toward high-energy radiation frequency can be due to the inverse turbulent cascade.

We cannot distinguish N-GRBs, S-GRBs, or H-GRBs by the low-energy photon index distributions. This indicates that the turbulent cascade does not significantly affect the low-energy index of the GRB prompt spectrum. However, it seems that S-GRBs have smaller $E_{\rm{peak}}$ values and H-GRBs have larger $E_{\rm{peak}}$ values. In other words, a harder LAT spectrum makes $E_{\rm{peak}}$ larger and a softer LAT spectrum makes $E_{\rm{peak}}$ smaller. This indicates that turbulent cascade takes effect on the GRB peak energy.

The numbers of N-GRBs, S-GRBs, and H-GRBs are comparable. This means that many GRBs have spectral evolution but some GRBs still have no strong spectral evolution. While some GRBs have a turbulent cascade process and some GRBs have an inverse turbulent cascade process. We consider both the LAT photon index sample obtained by the LAT observing time interval and that by the GBM observing time interval. Apparently, we do not find vital differences when we compare the above results derived from the two samples.

We finally summarize our main results mentioned in this subsection. Jitter radiation is a possible mechanism to explain GRB emission, and the power-law index of the jitter radiation is determined by the spectral index of the turbulence. Different turbulent cascade modes provide GRB spectral diversities shown in the different energy bands. We draw a cartoon to illustrate the GRB spectral diversities in Figure 9.

\section{Discussion}
We take the high-energy photon index of the Band function as the GBM photon index to compare to the LAT photon index.
The different fitting models to the GBM data were discussed \citep{gruber14,von14,yu16}. An exponential cutoff power-law fitting is also applied; a simple reason for this could be that the high-energy photons are not enough to constrain a simple power-law fitting. Corresponding to the radiative spectral property, the turbulent energy spectrum also has a cutoff power-law shape, and this shape was adopted in the jitter radiation \citep{mao07}.
\citet{gruber14} pointed out that the joint GBM and LAT data fitting for a GRB is helpful to investigate GRB spectral properties. An intrinsic spectral break in the high-energy band may happen on some GRBs. In this paper, we identify not only S-GRBs, but also N-GRBs and H-GRBs. The conclusions from the LAT photon index sample obtained by the LAT observing time interval and those by the GBM observing time interval do not have vital differences.

The electron pitch-angle distribution of the synchrotron radiation should be mentioned. In fact, this issue has been discussed in some of the literature \citep{es73a,es73b,ll00,ll02,baring04,yang18}. When the anisotropic pitch-angle is considered, the synchrotron radiation can reproduce a spectrum to fit the GRB prompt emission data.
IC and SSC mechanisms were also calculated in this case \citep{baring04}.
\citet{baring04} noted that the synchrotron with the small pitch-angle distribution has similar spectral shape to the jitter radiation presented by \citet{me00}. Here, we note two issues to further present the differences between the small pitch-angle synchrotron radiation and the jitter radiation.
First, the pitch-angle in the synchrotron radiation is defined as the angle between the electron momentum and the large scale magnetic field direction (e.g., Yang \& Zhang 2018). While the electron deflection angle smaller than the beaming angle is the condition for the jitter radiation, and the deflection angle is defined as the ratio between the perpendicular momentum of the Lorentz force and the electron momentum \citep{me00}. In principle, the pitch-angle and the deflection angle are different. However, if both angles are very small, they are quantitatively the same. Second, synchrotron radiation takes place in the large-scale bipolar magnetic field, and electrons have helical orbits, while jitter radiation occurs in the random and small-scale magnetic field, and the electrons have a ``jitter" trajectory due to the random Lorentz force by the magnetic elements with the same strength. Although the small pitch-angle of the synchrotron and the small deflection angle of the jitter radiation are similar, the small pitch-angle synchrotron radiation and the jitter radiation are two different physical mechanisms.

Although we investigate the GRB properties in the energy band above $E_{\rm{peak}}$, jitter radiation can produce a general spectral shape to compare to observational data. Here, we take the results from \citet{me00} as an example. In the low-energy band, the shape is related to the electron deflection angle compared to the beaming angle. The spectral properties in the low-energy band described by the Band function can be explained by the jitter radiation.
When we consider the hydrodynamic/MHD effect, we expect that stronger turbulence has a positive effect, because random and small-scale magnetic field is induced by the turbulence in our scenario.
Moreover, we note that the peak energy produced by the jitter radiation is larger than that produced by the synchrotron radiation. This was also mentioned by \citet{kelner13}. Thus, stronger turbulence makes larger $E_{\rm{peak}}$ number. In this paper, we identify that the turbulent cascade makes the spectral evolution above $E_{\rm{peak}}$, and the $E_{\rm{peak}}$ values are also affected by the turbulent cascade as shown in Figures 7 and 8.

Synchrotron radiation has a power-law spectrum with a spectral index $(n-1)/2$, where $n$ is the index of the power-law electron energy distribution. The spectral index of the GRB prompt emission above $E_{\rm{peak}}$ follows the term of $(n-1)/2$ \citep{mes93,mesetal94}. \citet{spada00} confirmed the slope in the case that the radiative frequency is larger than the synchrotron peak frequency but smaller than the cooling frequency. If the radiative frequency is larger than the cooling frequency, the slope is modified to be $n/2$.
In principle, GRB spectral data can be well explained by the synchrotron radiation with $n=2.2-2.5$ \citep{band93,preece98}.
In general, when we consider a power-law distribution with a slope $\beta$ for the initial photon field,
the IC/SSC power-law spectral index is $\beta-1$.
Therefore, the IC/SSC mechanism usually produces a harder spectrum in the GeV-GRB case \citep{sari01}.
GRB 090510 and GRB 130427A are two examples that the harder spectra detected by LAT were identified \citep{corsi10,fan13,liu13,tam13}. However, we note other interpretations. The LAT-detected emission of GRB 090510 can be explained by the synchrotron radiation from the forward shock, and the peak of the LAT-detected emission can be explained by the SSC process from the reverse shock \citep{fra16a}. When LAT-detected emission is treated by GRB afterglow, the density numbers of the interstellar medium adopted by \citet{corsi10} and \citet{joshi19} are not fully consistent. For GRB 130427A, the multiwavelength (LAT-band, X-ray, and optical) lightcurves can be explained by the synchrotron radiation \citep{kou13,perley14,fra16b}, while the synchrotron radiation was not supported by the Swift X-ray observation \citep{de16}.
Besides the issues mentioned above, the IC/SSC mechanism seems difficult to use to interpret the softer spectra in LAT GRBs, although the harder spectra in LAT GRBs can generally be explained by IC/SSC mechanism.

It is more complicated when we consider some recent works on the particle acceleration related to the microdynamics. The particle acceleration due to the kinetic turbulence and the magnetic reconnection sometimes has a hard electron energy spectrum with $n<2.0$ \citep{na15,werner16,zhdankin17,pe18}. Therefore, it is quite complex to explain the GRB spectral diversities in the high-energy bands by the IC/SSC mechanism. A recent work on the plasma turbulence in the magnetized case can produce the hard synchrotron spectra \citep{comisso20}, and it could also be helpful to understand the S-GRBs presented in this paper. Furthermore, when we consider the turbulence and the related magnetic reconnection to have particle acceleration, electron cooling should be considered as well. In principle, the jitter radiation index is only related to the spectral index of the turbulence in our scenario, and it is not related to the electron energy distribution. However, the radiation depends on the acceleration and cooling that have different results in different energy bands. It is possible that the cooling frequency between GBM and LAT energy bands makes the GRB spectral diversities presented in this paper.

The extra power-law added on the Band function has been considered in some spectra listed in the first LAT GRB catalog \citep{ack13}. We think that this is suitable for the harder spectra toward the higher radiation frequency. H-GRBs classified in this paper can be identified by the spectral fitting with the Band function plus a power-law in future samples. Although the IC/SSC mechanism can be one possible explanation to H-GRBs, we consider inverse turbulent cascade to explain these H-GRBs. While the IC/SSC mechanism is difficult to use to explain S-GRBs, we think that turbulent cascade makes GRBs softer toward high-energy radiation frequency. Moreover, an exponential cutoff can be added to the Band function. In this paper, we describe the turbulent energy spectrum by a simple power law. Sometimes, a turbulent energy spectrum can be described by a power law with an exponential cutoff, and the cutoff indicates that the turbulent cascade at small lengthscales is terminated. Thus, this hydrodynamic feature makes a cutoff feature in the radiative spectrum. A joint data analysis of GBM and LAT can be helpful to further investigate GRB spectral properties.

The turbulence working on the GRB emission was suggested in the early time by \citet{narayan09}, \citet{kumar09}, and \citet{lazar09}. \citet{lemoine13} introduced the decaying microturbulence into the relativistic blast wave, and the GRB spectrum produced by the synchrotron radiation can be modified. Because GRB spikes are emitted in very small timescales, the lengthscale that corresponding to the timescale should also be very small. When we perform the jitter radiation to study the GRB emission, we consider the turbulence cascade process in this paper. Thus, the GRB spectral diversity in the high-energy bands can be explained. However, when we consider the inverse turbulent cascade \citep{zrake14,bran15}, the harder turbulent spectra are not universal \citep{zrak16}. The combination of the cascade and the inverse cascade mentioned by \citet{franci17} should be further investigated in order to explore which process, cascade, or inverse cascade, is dominated in the GRB case.
The turbulent energy spectrum under the turbulent stress was examined recently \citep{bran19}. In fact, turbulent cascade shows quite different features as the energy injection inputs at different lengthscales \citep{ale18}.
Therefore, we need to consider more complicated cases to further perform a quantitatively analysis to the observed GRB spectra.
In the meantime, we cannot simply rule out the IC/SSC mechanism. For instance, the electrons producing jitter photons can be scattered by the jitter photons themselves, and the scattered photons can be shown in the GeV energy band. We have called this mechanism the jitter self-Compton (JSC; Mao \& Wang 2012) process. Similar to the IC/SSC mechanism, JSC mechanism can be adopted to explain harder spectra in H-GRBs. However, the JSC mechanism is still difficult for the explanation of the soft spectra in S-GRBs.

When GRB outflow is magnetic-dominated, electron cooling should be very fast. Some physical mechanisms for the magnetic energy dissipation are necessary. Here, we emphasize that magnetic reconnection can be a natural process for the energy dissipation. For example, \citet{mc12} presented a general process of magnetic reconnection in GRB outflows. A recent work on the turbulent reconnection for GRBs was presented \citep{la19}. Here, we mention the relation between magnetic reconnection and turbulence. Magnetic reconnection can drive turbulence, and vice versa. Because the magnetic field is random and small scale in the jitter radiation framework, we infer that magnetic reconnection and turbulence take place at the plasma lengthscales. In the meantime, synchrotron radiation and the IC/SSC process associated with the magnetic reconnection have been discussed (e.g., Uzdensky 2016).

We note that some special properties are shown in two specific GRBs. GRB 100724B detected by LAT has a photon index of 4. This burst has a thermal component shown in the GBM spectrum \citep{guiriec11}. However, it seems that the spectral shape in the high-energy bands can be fitted by a simple power law. Thus, the LAT-detected photon index is reliable. The GRB 170405A extended emission detected by LAT has a time evolution with its spectra from the soft to hard state \citep{ari20}. We should pay attention to more complicated physical issues when we consider case studies.

The turbulent cascade development from large to small lengthscales takes time. The cascade process depends on the detailed physical issues. Here, we adopt the turbulence induced by the magnetic reconnection to be an example \citep{lo17}. We take the turbulent eddy turnover time in the magnetic reconnection scenario as the cascade timescale. The eddy turnover time is estimated as $t_{\rm{eddy}}=(\lambda L)^{1/2}/V_A$, where $\lambda$ is the turbulent eddy lengthscale, $L$ is the outer scale of the turbulence, and $V_A\sim c$ is the Alfv\'{e}n speed. We take the width of the shock $R/\Gamma^2$ as the outer lengthscale of the turbulence, and the lengthscale of the turbulent eddy is $R/\Gamma\gamma_t$, where $R$ is the GRB fireball radius, $\Gamma$ is the bulk Lorentz factor of the fireball shock, and $\gamma_t$ is the turbulent Lorentz factor. Thus, we obtain
\begin{equation}
t_{\rm{eddy}}=10.0(\frac{R}{1.0\times 10^{15}~\rm{cm}})(\frac{\Gamma}{100.0})^{-3/2}(\frac{\gamma_t}{10.0})^{-1/2}~\rm{s}.
\end{equation}
This means that the turbulence needs about 10.0 s to have the energy cascade in the GRB case when we consider the fireball radius at $10^{15}$ cm.
We note that the observed timescale should be the intrinsic timescale divided by the bulk Lorentz factor $\Gamma$ due to the compression in the observer time caused by the emitting shell moving toward the observer \citep{piran04}. This yields a number of 0.1 s if we set $\Gamma=100.0$.
The LAT GRBs usually have a time delay compared to the GBM GRBs. We may consider the turbulent eddy turnover time as the delay time due to the energy cascade.
The LAT GRBs with the time delay of less than 1 s can be explained in this paper. However, it is difficult to explain some LAT GRBs with the time delay larger than 1 s \citep{ajello19}. Moreover, the case of GRB 130427A is an exception, and its LAT emission was detected earlier than the GBM emission \citep{preece14}.
The decay time can be also related to the acceleration and cooling of the electrons as presented in the former discussion, and the cooling may induce the GRB spectral evolution \citep{mo16}. We may further consider the time evolution of the turbulent cascade in the future.

\section{Conclusions}
We illustrate that the GRB spectra obtained by the Fermi-GBM and Fermi-LAT observations have diversities. We propose the turbulent cascade to explain the GRB spectral properties in the jitter radiation framework. Because the jitter emission at higher radiative frequency is produced at smaller lengthscale and the radiation spectral index is determined by the spectral index of the turbulence, the turbulent spectral modification due to the turbulent cascade makes the GRB spectral diversities in the high-energy bands. We need a large data sample for statistical analysis and we should consider more complicated physical issues to do case studies in the future.

\acknowledgments
We are grateful to the referee for a careful review and very helpful suggestions.
J.M. is supported by the National Natural Science Foundation of China 11673062, the Hundred Talent Program of Chinese Academy of Sciences, and the Oversea Talent Program of Yunnan Province. J.W. is supported by the National Natural Science Foundation of China (11573060 and 11661161010).

\clearpage




\clearpage
\begin{figure}
\center
\includegraphics[scale=0.55]{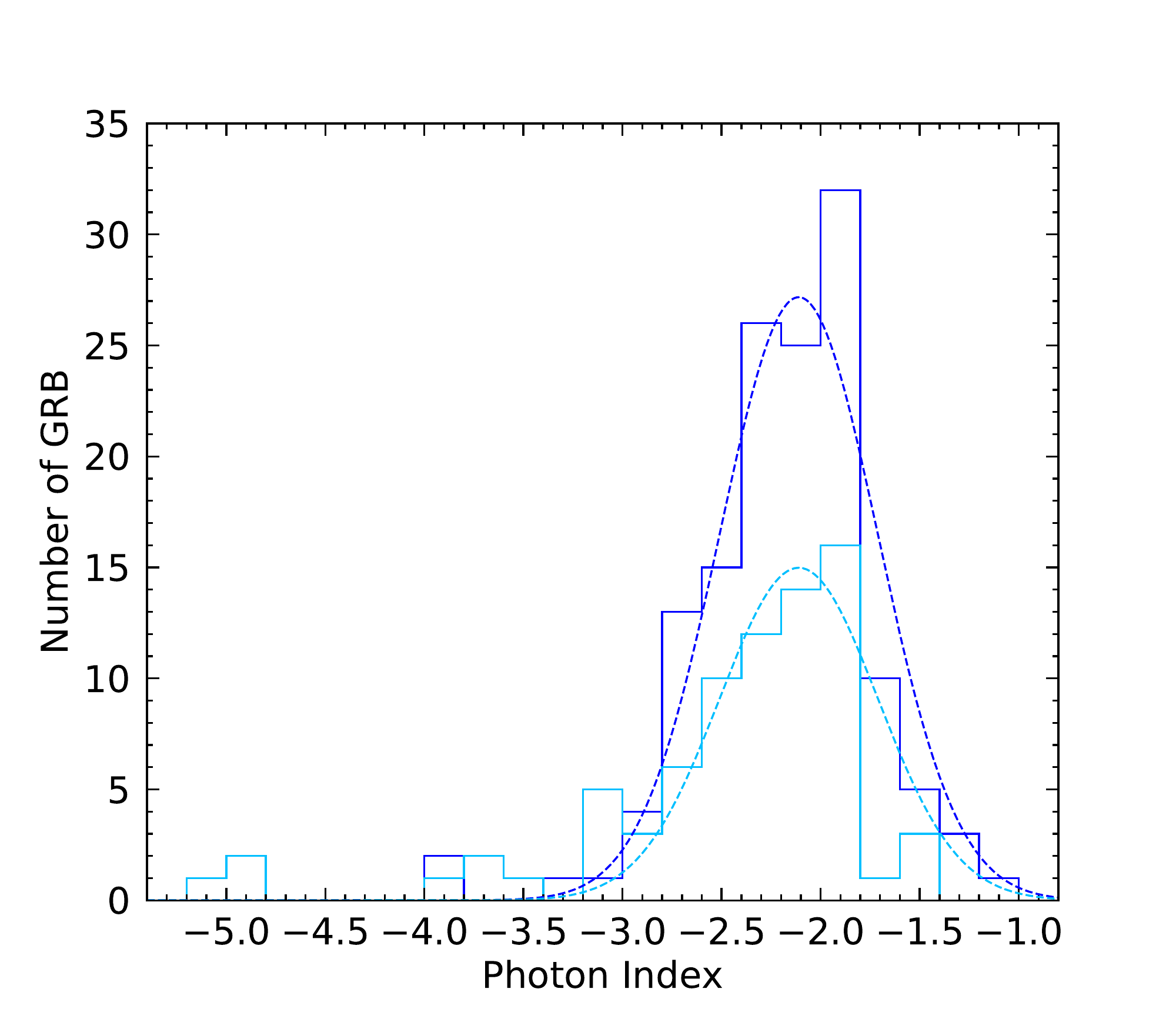}
\includegraphics[scale=0.55]{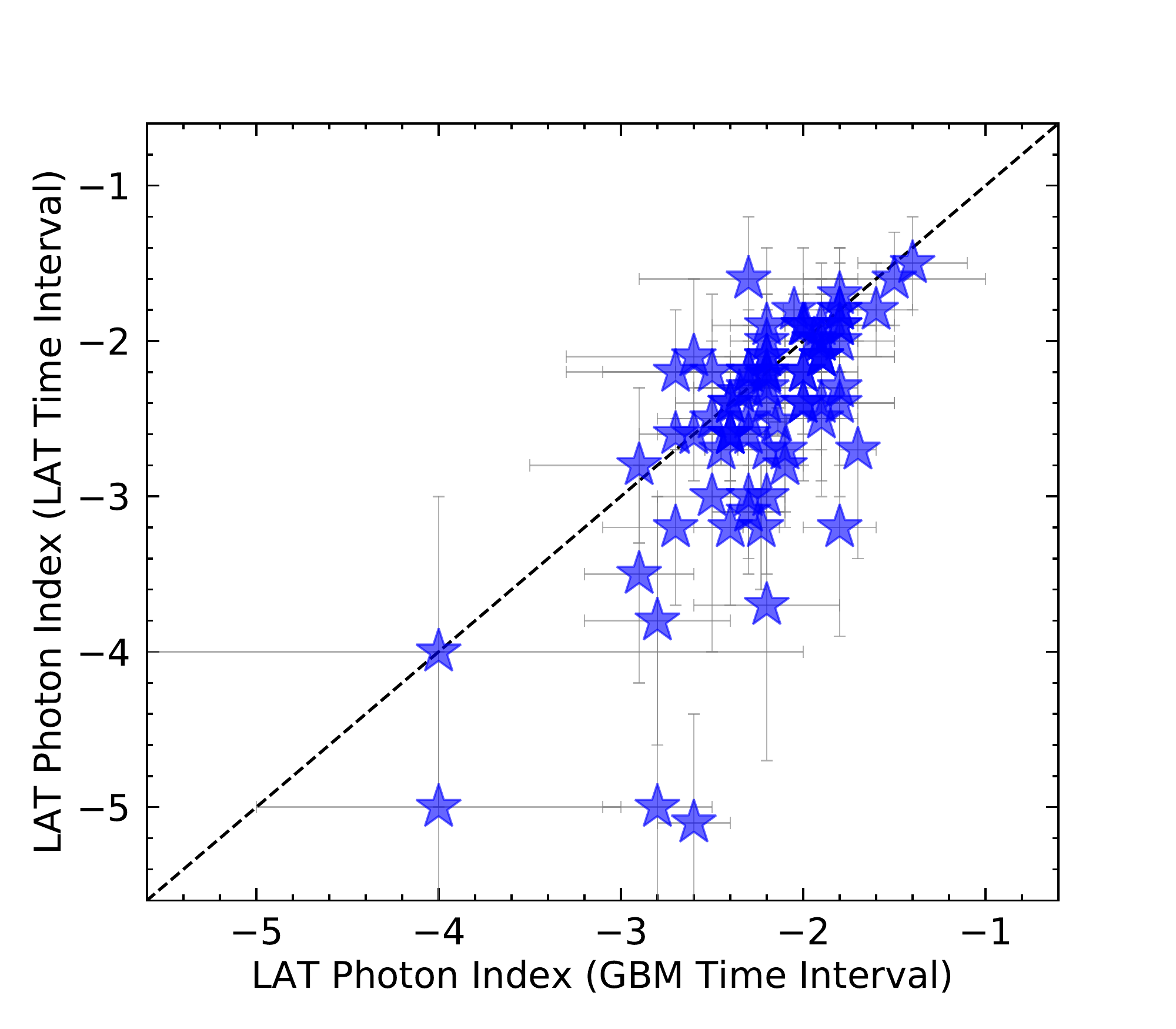}
\caption{\footnotesize{Upper panel: LAT photon index distribution with a Gaussian fitting. The curves with the dark blue color indicate the data from the LAT observing time interval, and the curves with the faint blue color indicate the data from the GBM observing time interval. Lower panel: the LAT photon index obtained in the LAT observing time interval vs¡£ the LAT photon index obtained in the GBM observing time interval. The dashed line indicates that two photon indices are equal.}
\label{fig1}}
\end{figure}

\begin{figure}
\center
\includegraphics[scale=0.4]{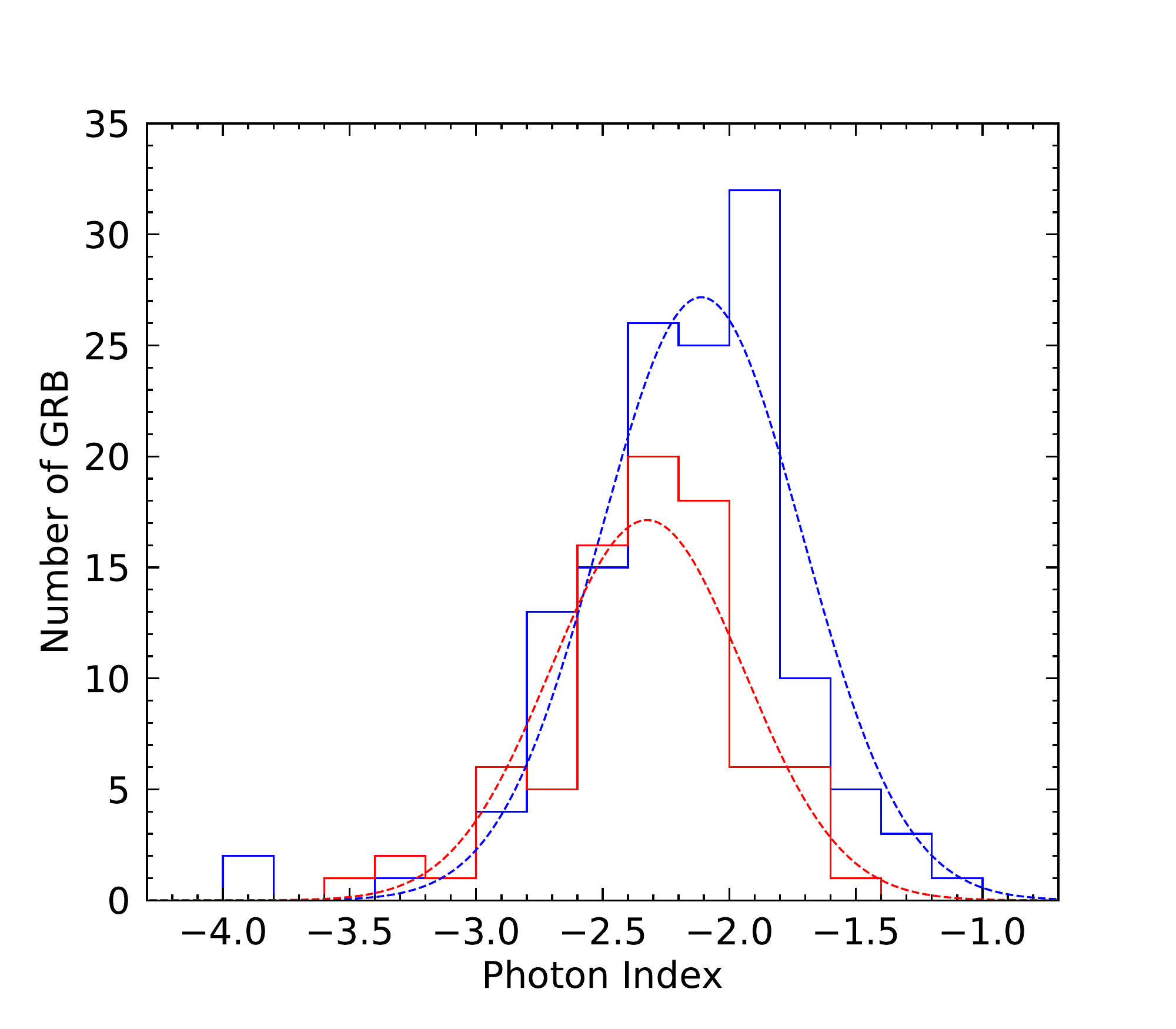}
\includegraphics[scale=0.4]{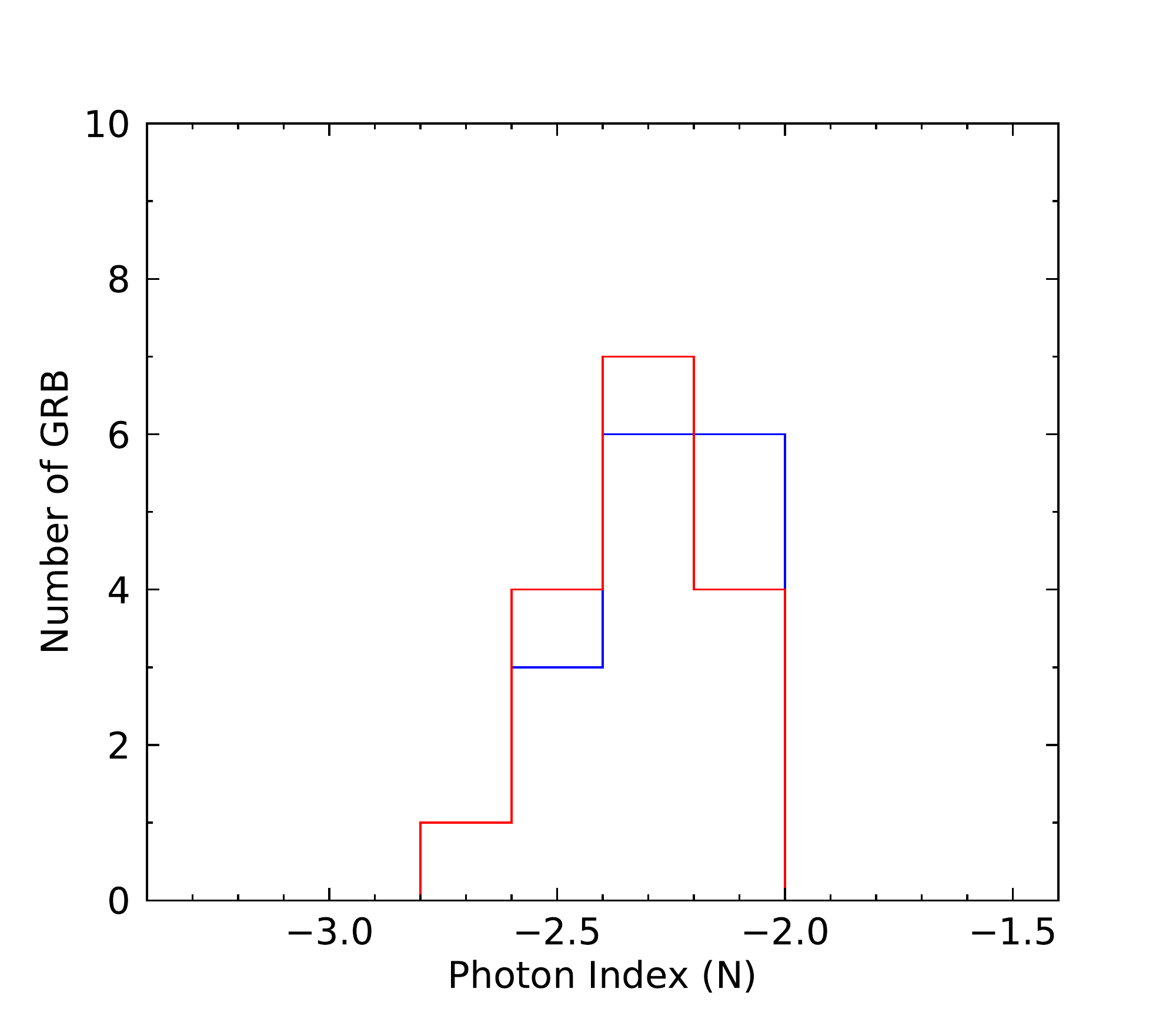}
\includegraphics[scale=0.4]{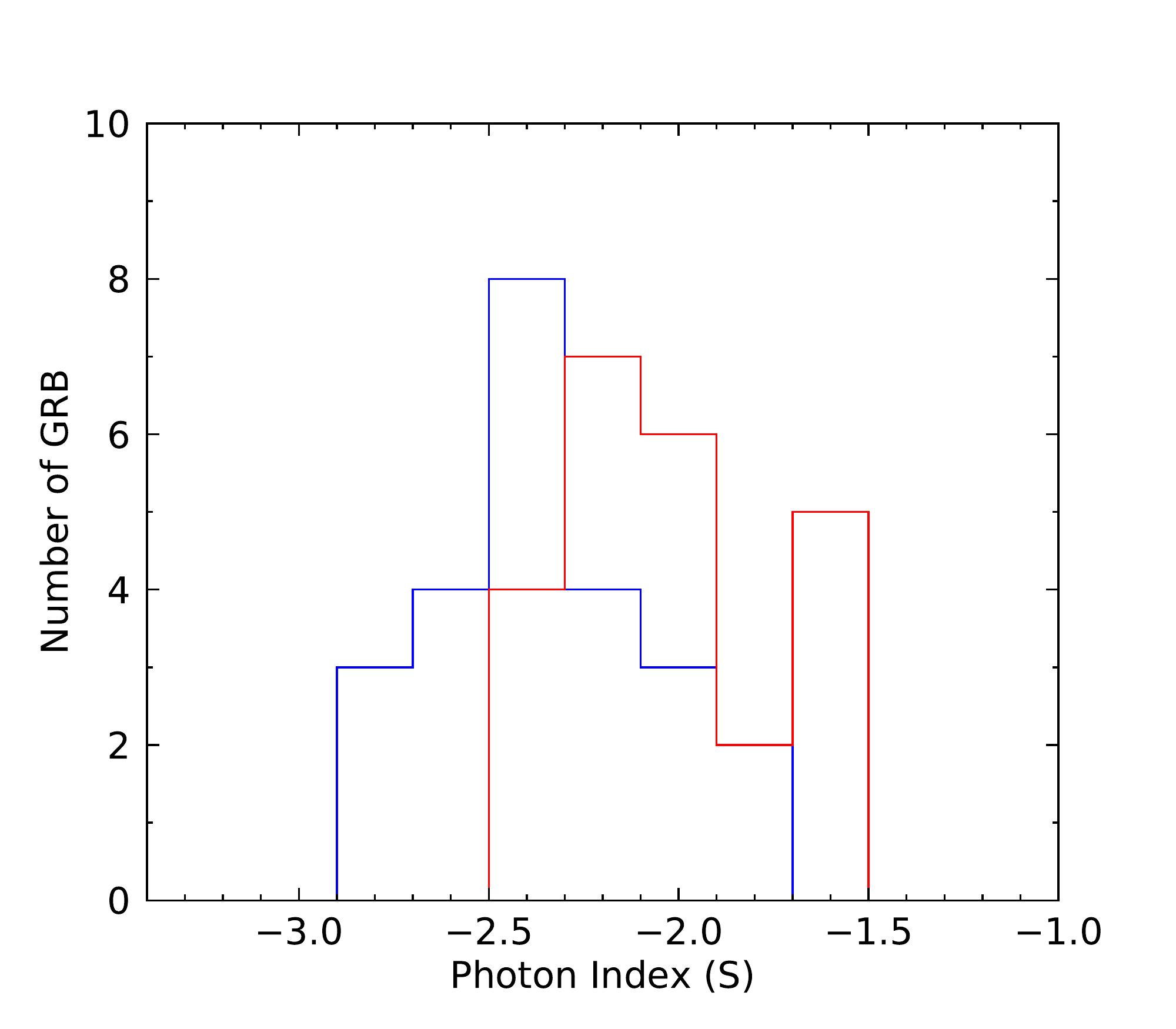}
\includegraphics[scale=0.4]{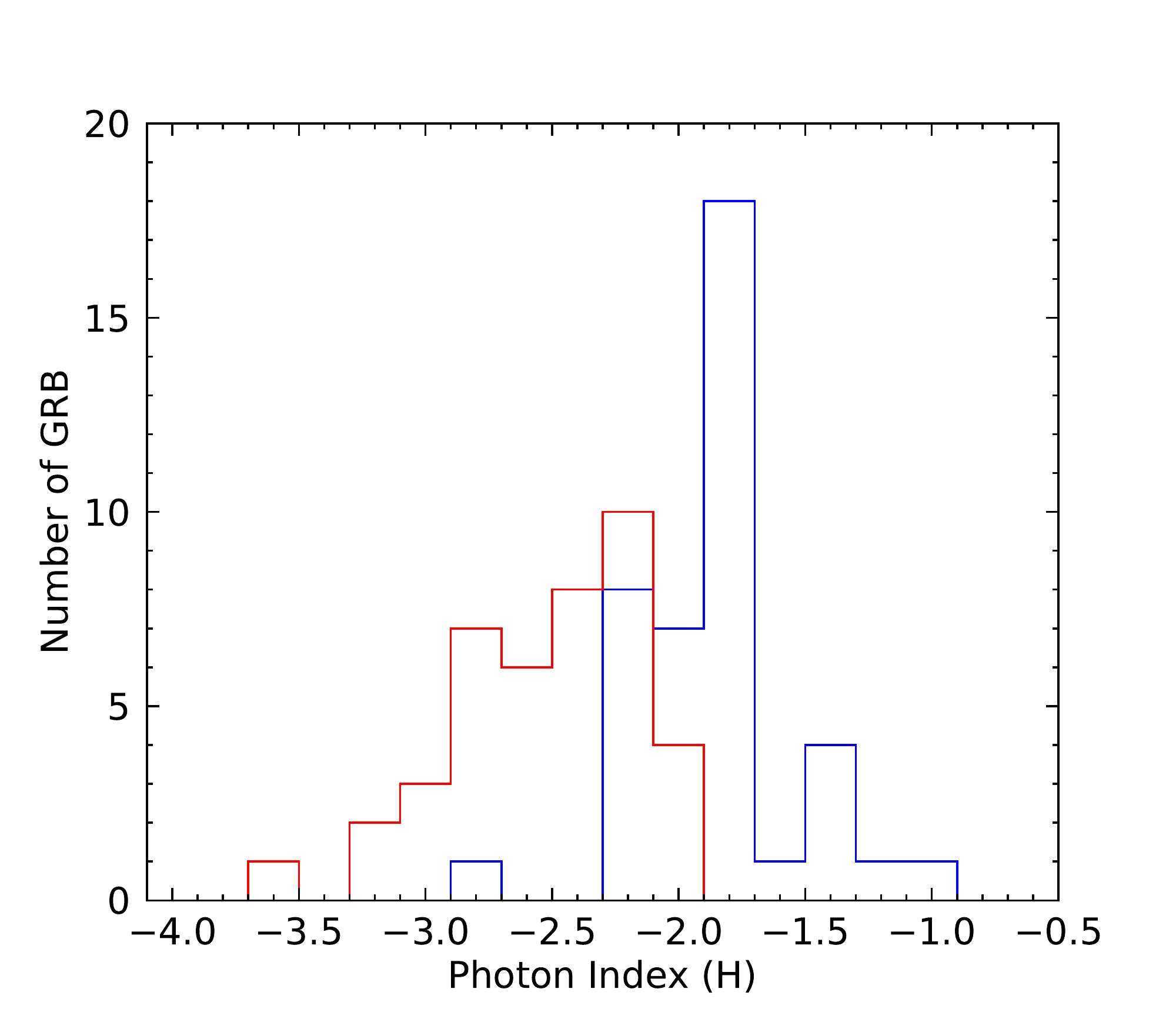}
\caption{GRB high-energy photon index distribution. The distributions with blue and red colors indicate the LAT and GBM samples, respectively. The LAT photon indices are obtained from the LAT observing time interval.
Upper left panel: all the GRB photon indices obtained from either LAT or GBM. The Gaussian fittings are also shown. Upper right panel: N-GRB photon index distribution. Lower left panel: S-GRB photon index distribution. Lower right panel: H-GRB photon index distribution.
\label{fig2}}
\end{figure}

\begin{figure}
\center
\includegraphics[scale=0.4]{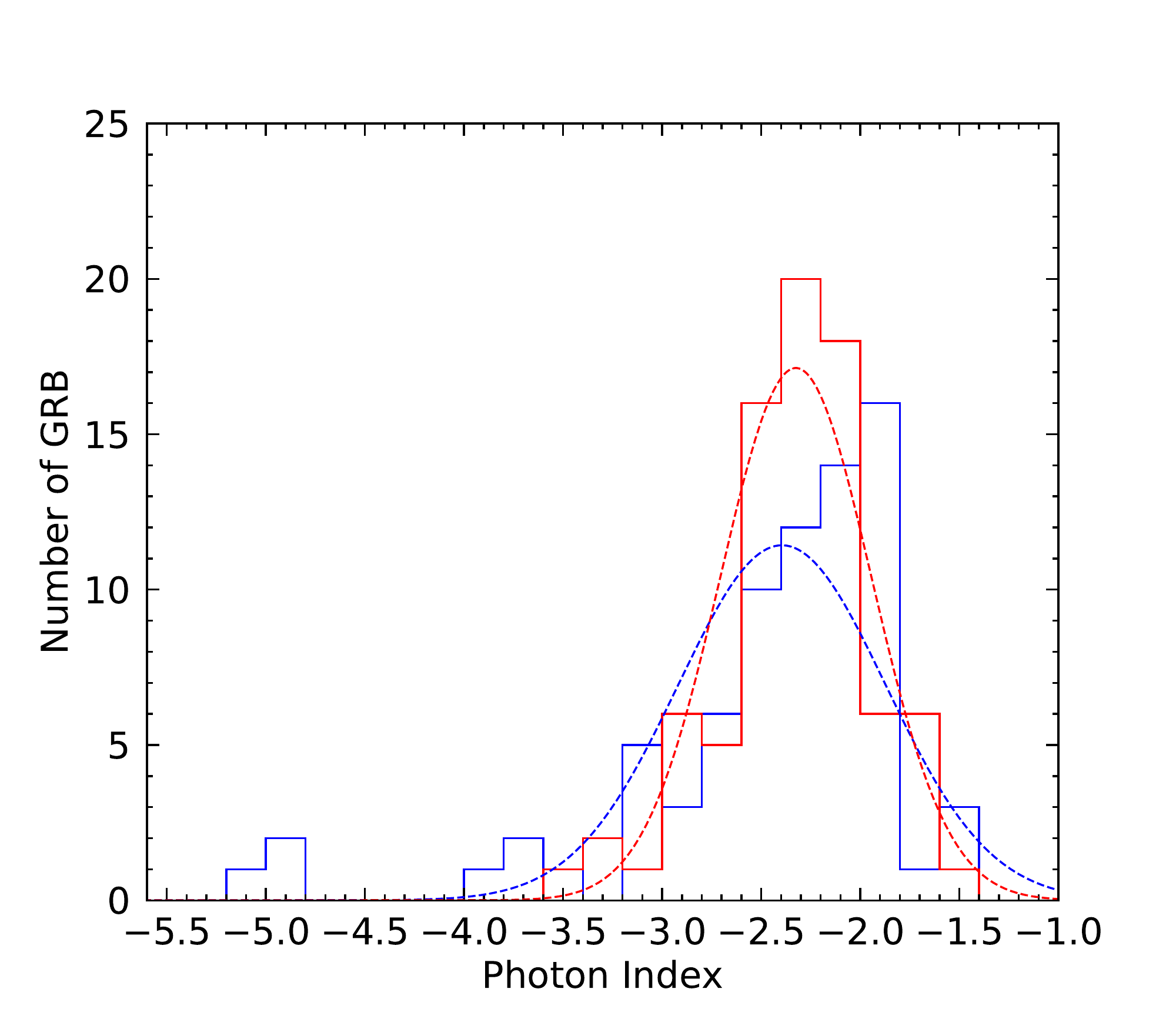}
\includegraphics[scale=0.4]{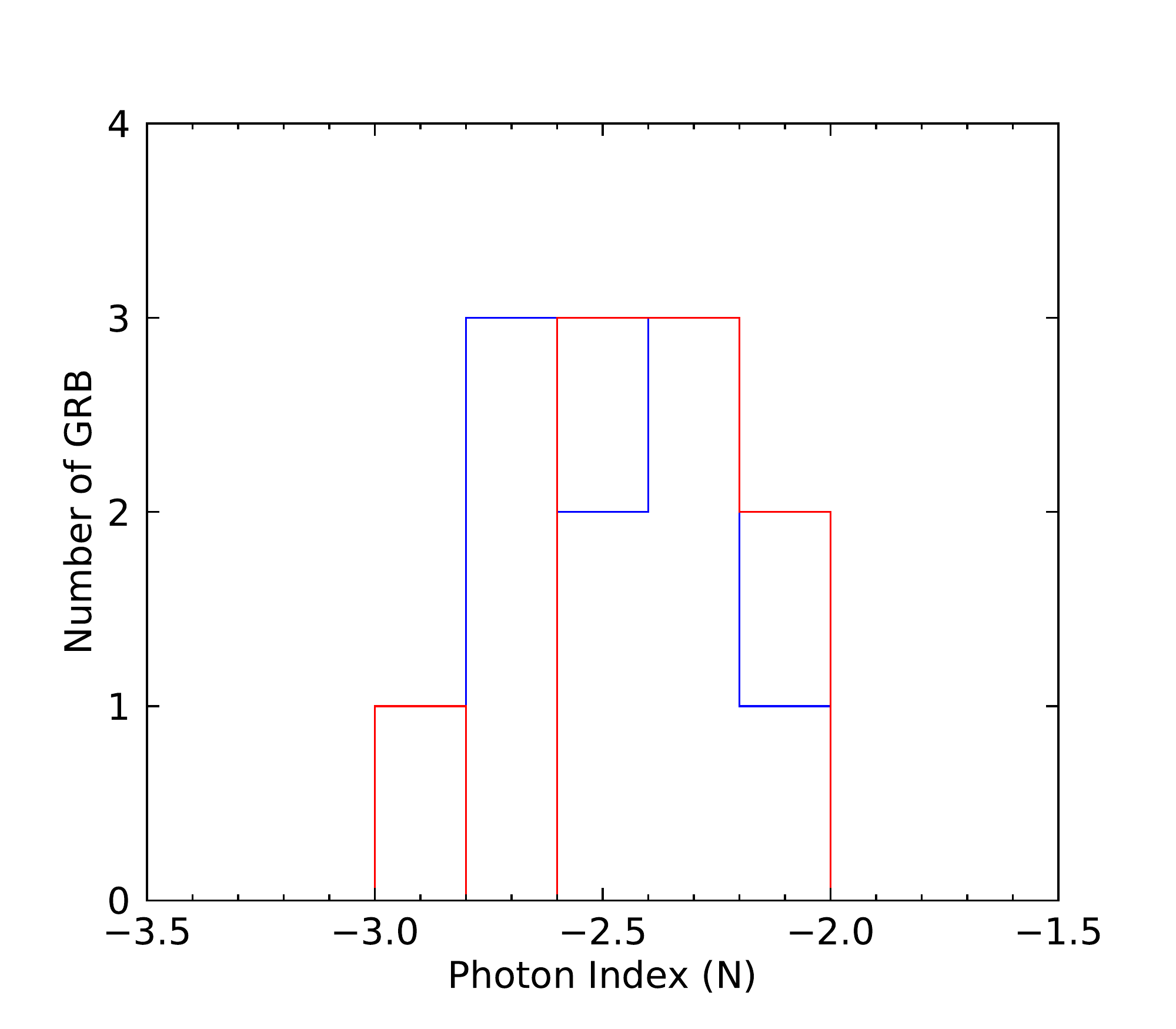}
\includegraphics[scale=0.4]{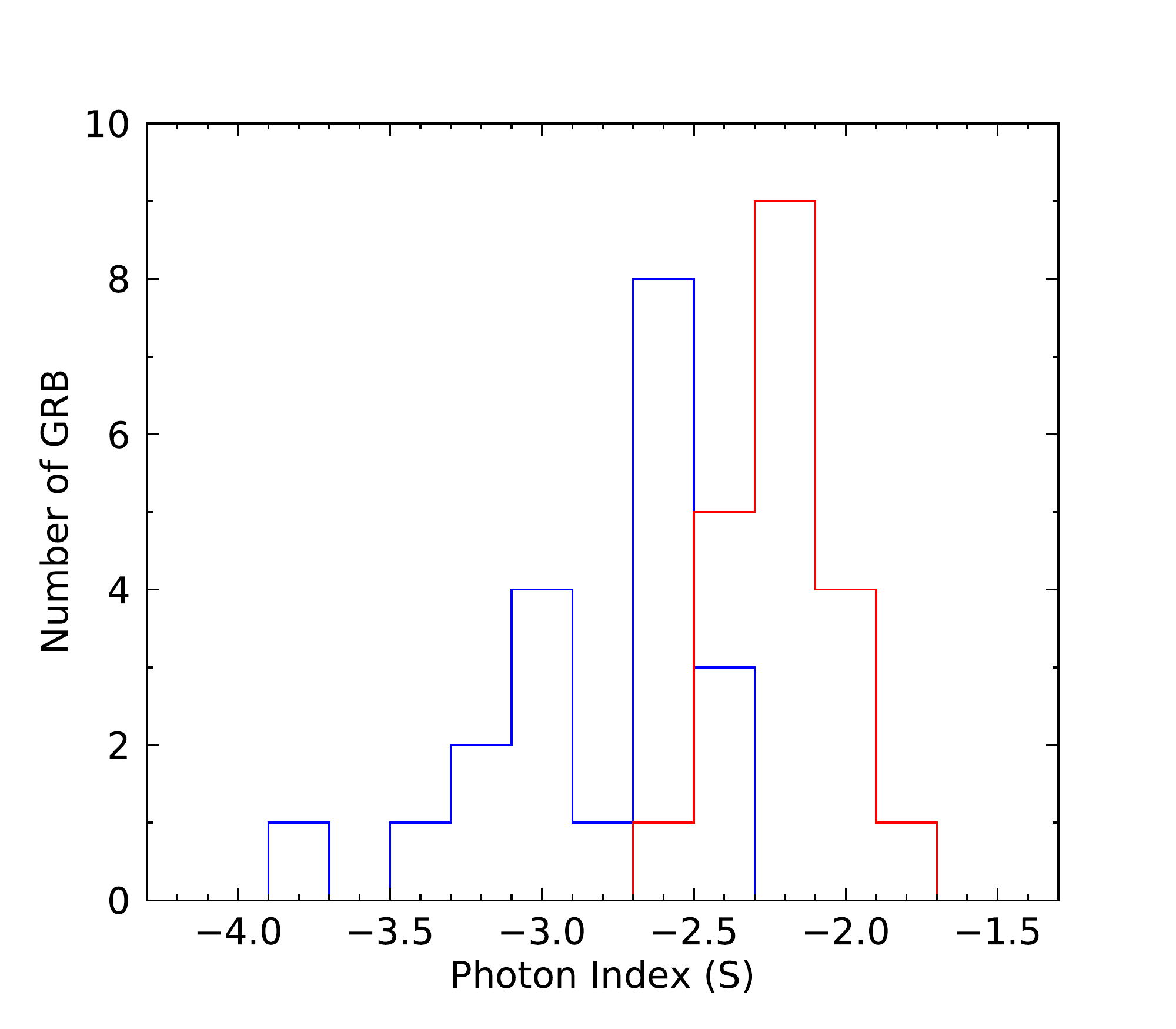}
\includegraphics[scale=0.4]{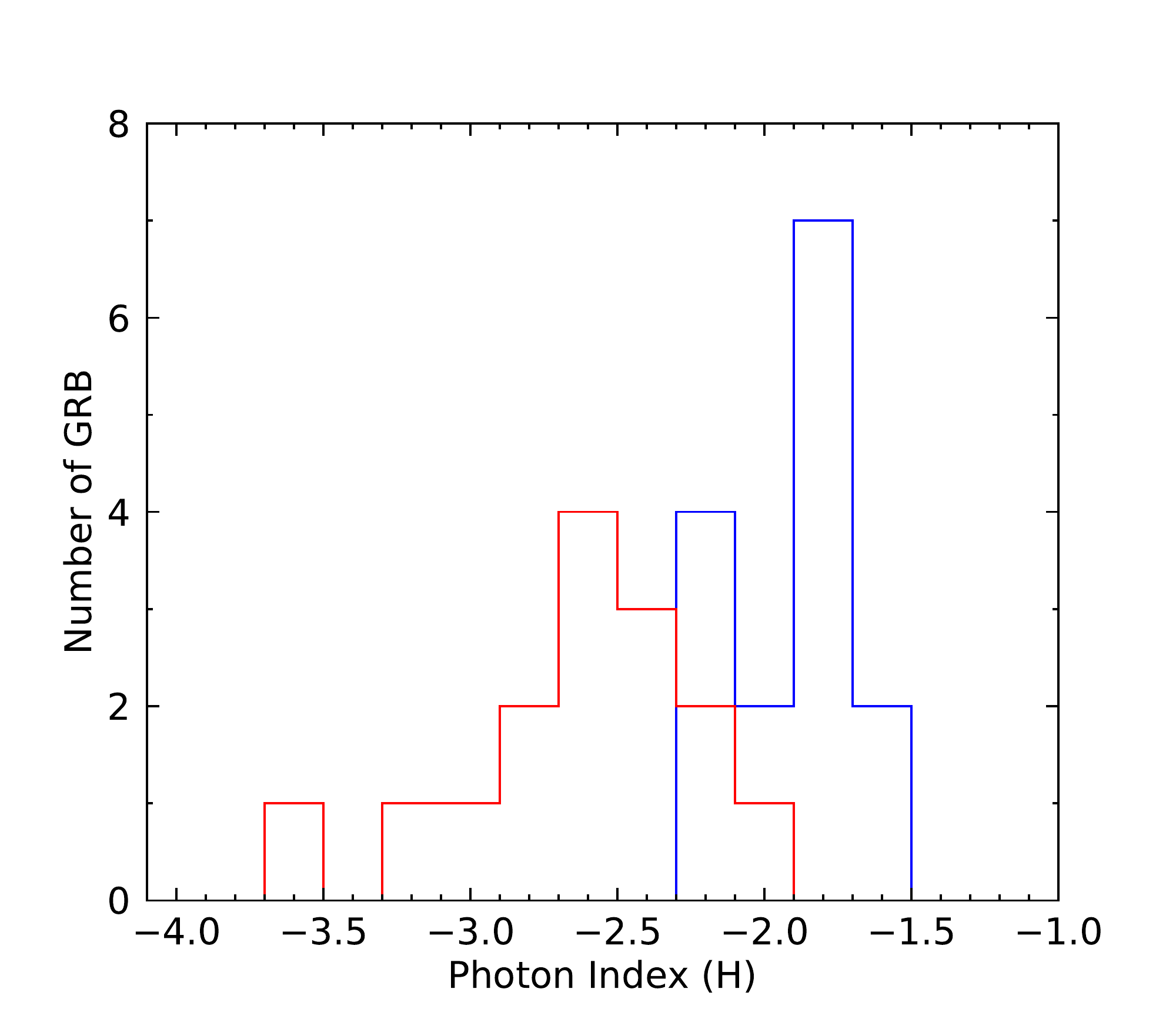}
\caption{GRB high-energy photon index distribution. All notations are the same as those in Figure 2, except that the LAT photon indices are obtained from the GBM observing time interval.
\label{fig3}}
\end{figure}

\begin{figure}
\center
\includegraphics[scale=0.55]{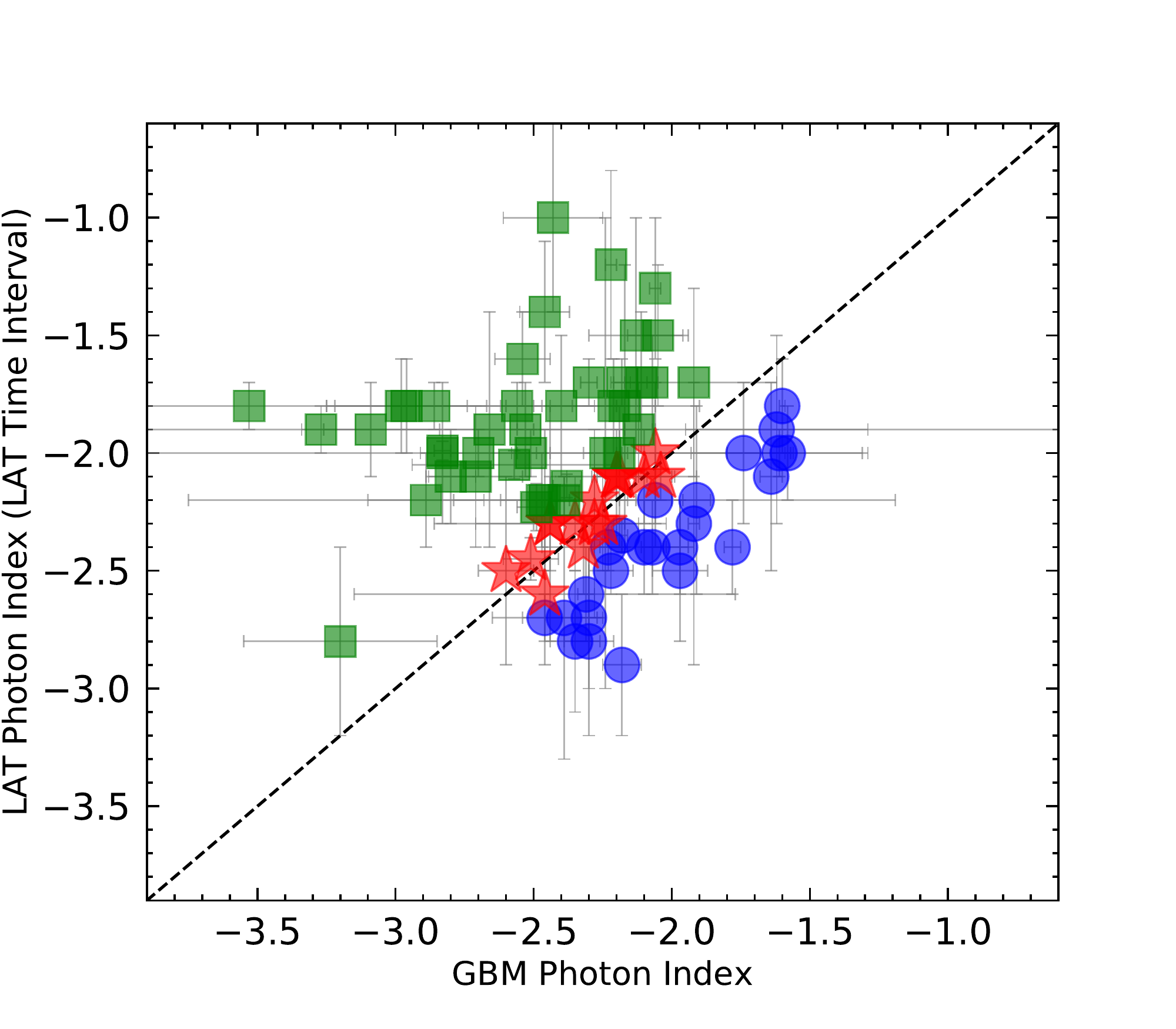}
\includegraphics[scale=0.55]{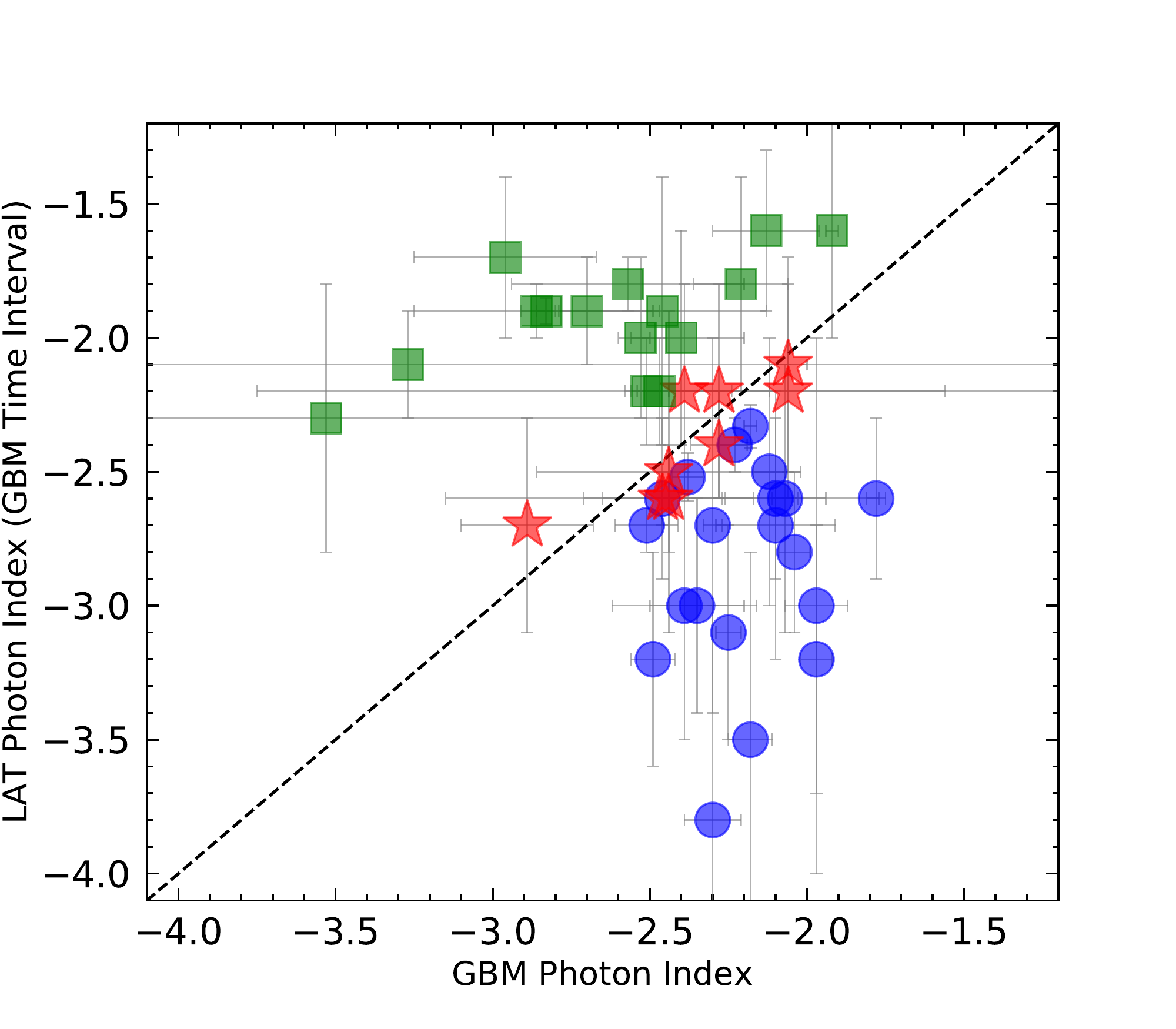}
\caption{\footnotesize{LAT photon index vs. GBM photon index. Red stars indicate N-GRBs, green squares indicate H-GRBs, and blue dots indicate S-GRBs. The dashed line indicates that the LAT photon index is equal to the GBM photon index. Upper panel: the LAT photon indices obtained from the LAT observing time interval. GRB 100724B with the LAT photon index $-4.00\pm 1.00$ is not shown in the plot. Lower panel: the LAT photon indices obtained from the GBM observing time interval. GRB 100724B with the LAT photon index $-5.00\pm 1.00$, GRB 160821A with the LAT photon index $-5.10\pm 0.70$, and GRB 170405A with the LAT photon index $-5.00\pm 2.00$ are not shown in the plot.}
\label{fig4}}
\end{figure}

\begin{figure}
\center
\includegraphics[scale=0.4]{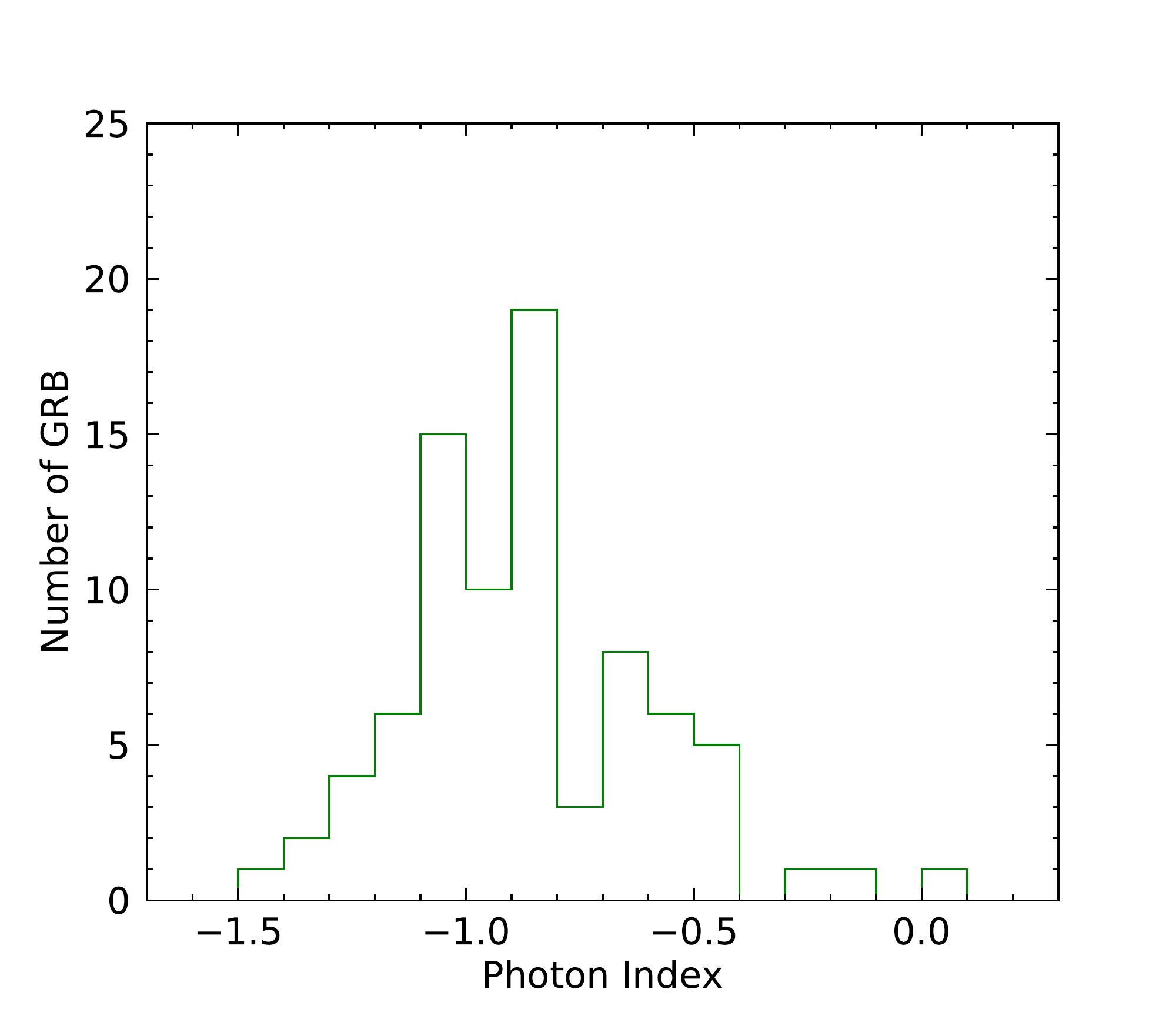}
\includegraphics[scale=0.4]{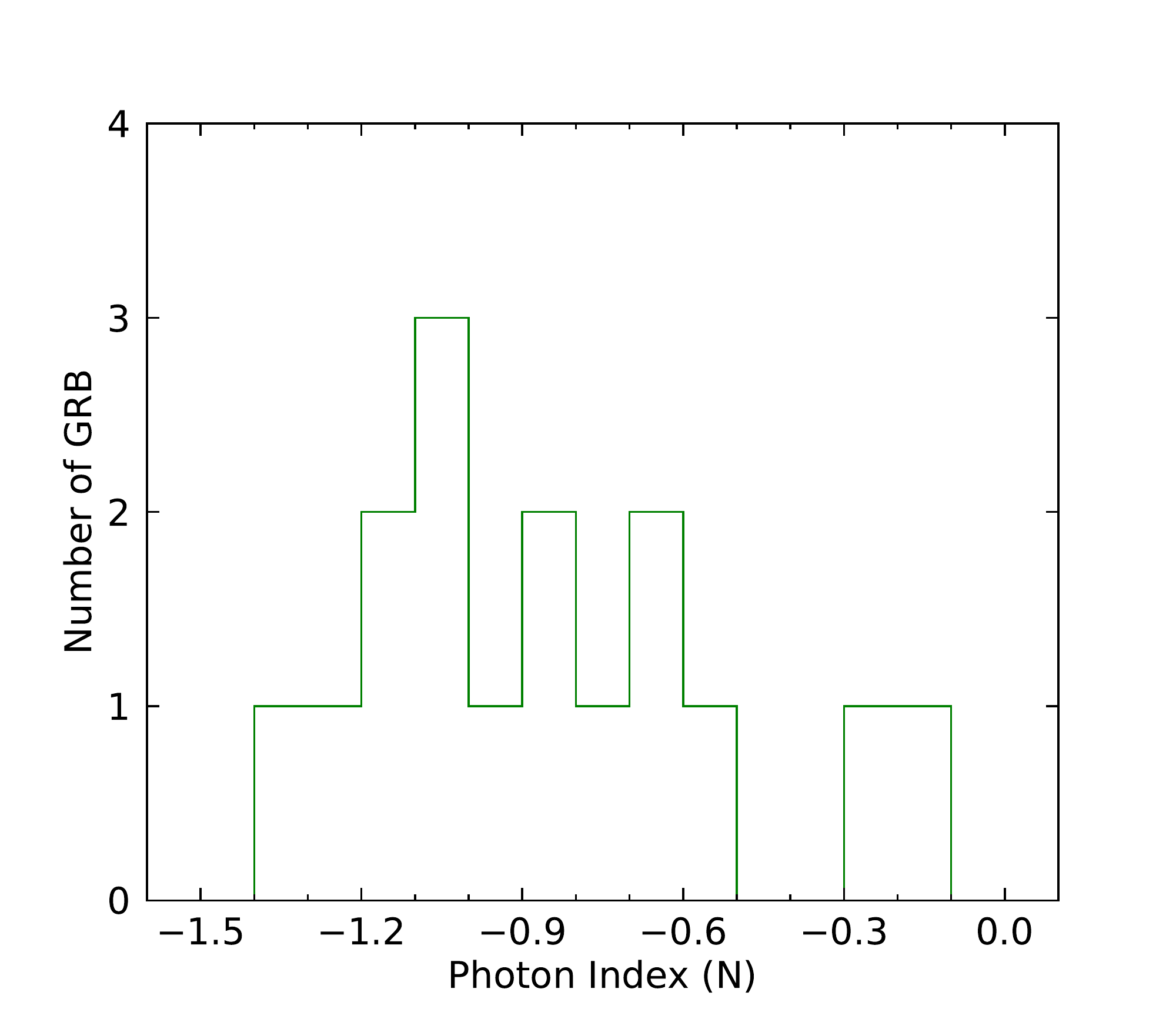}
\includegraphics[scale=0.4]{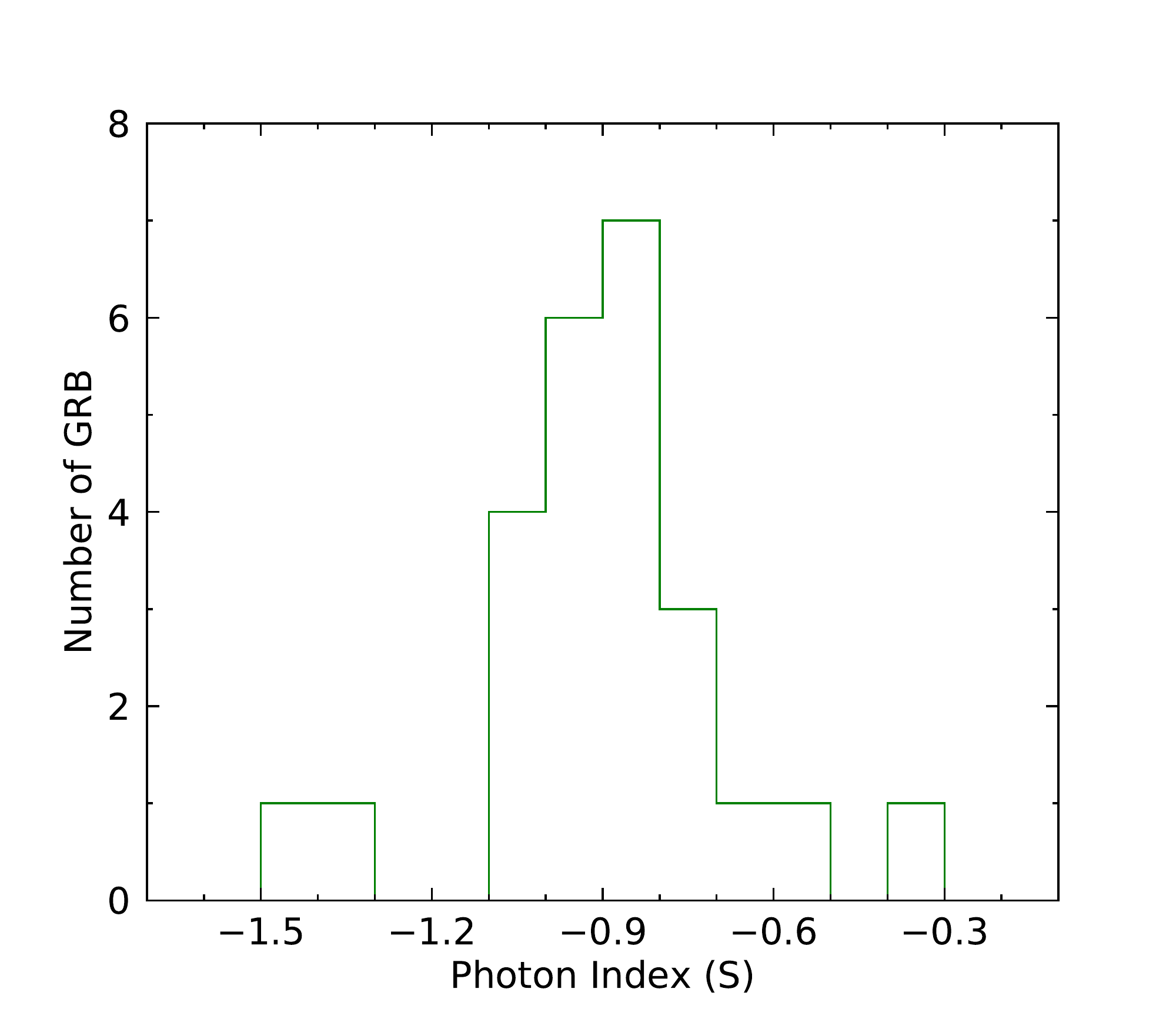}
\includegraphics[scale=0.4]{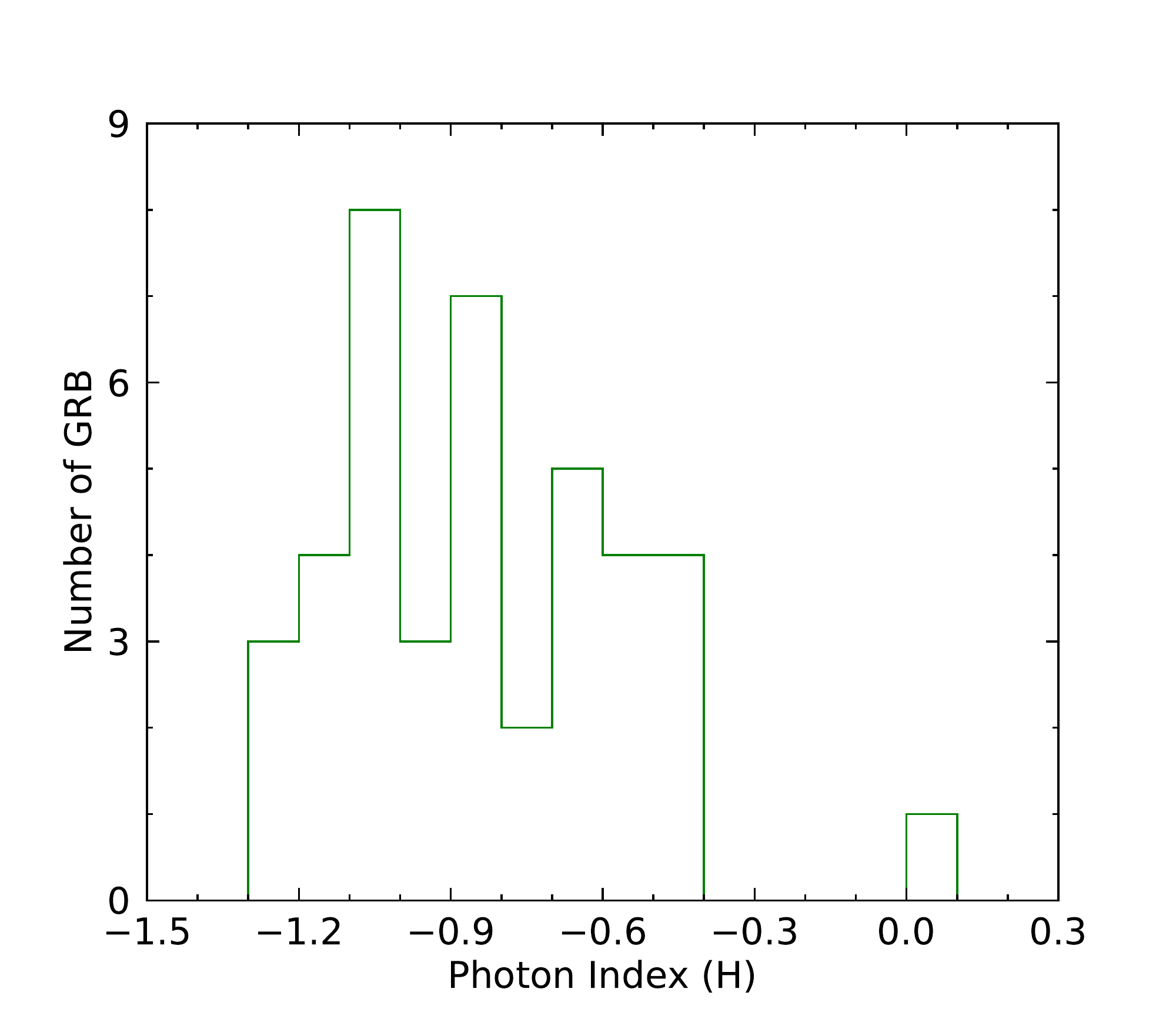}
\caption{GRB low-energy photon index distribution. The GRBs are selected from those with the LAT photon indices obtained from the LAT observing time interval. Upper left panel: all the GRB photon index distributions. Upper right panel: N-GRB photon index distribution. Lower left panel: S-GRB photon index distribution. Lower right panel: H-GRB photon index distribution.
\label{fig5}}
\end{figure}

\begin{figure}
\center
\includegraphics[scale=0.4]{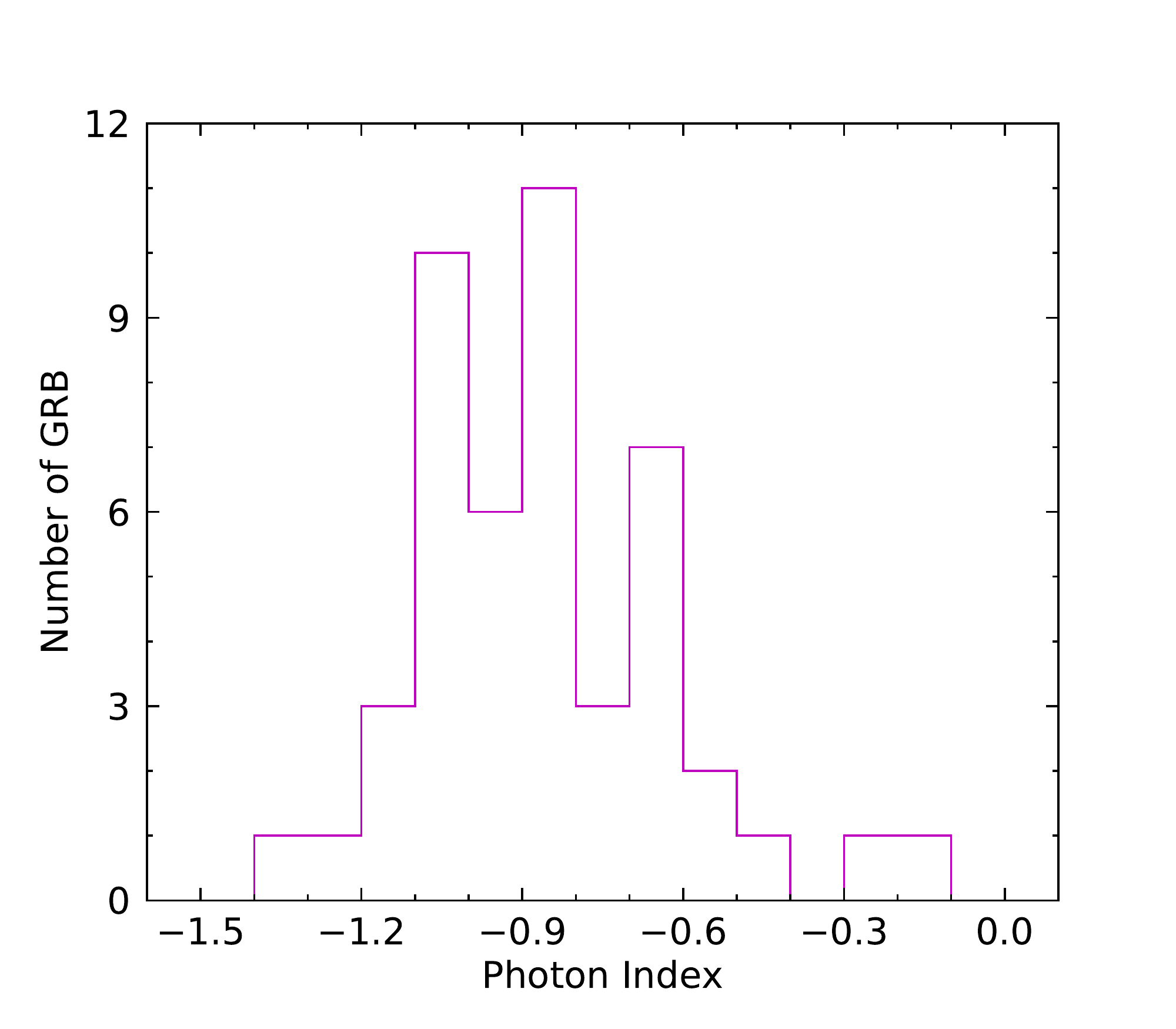}
\includegraphics[scale=0.4]{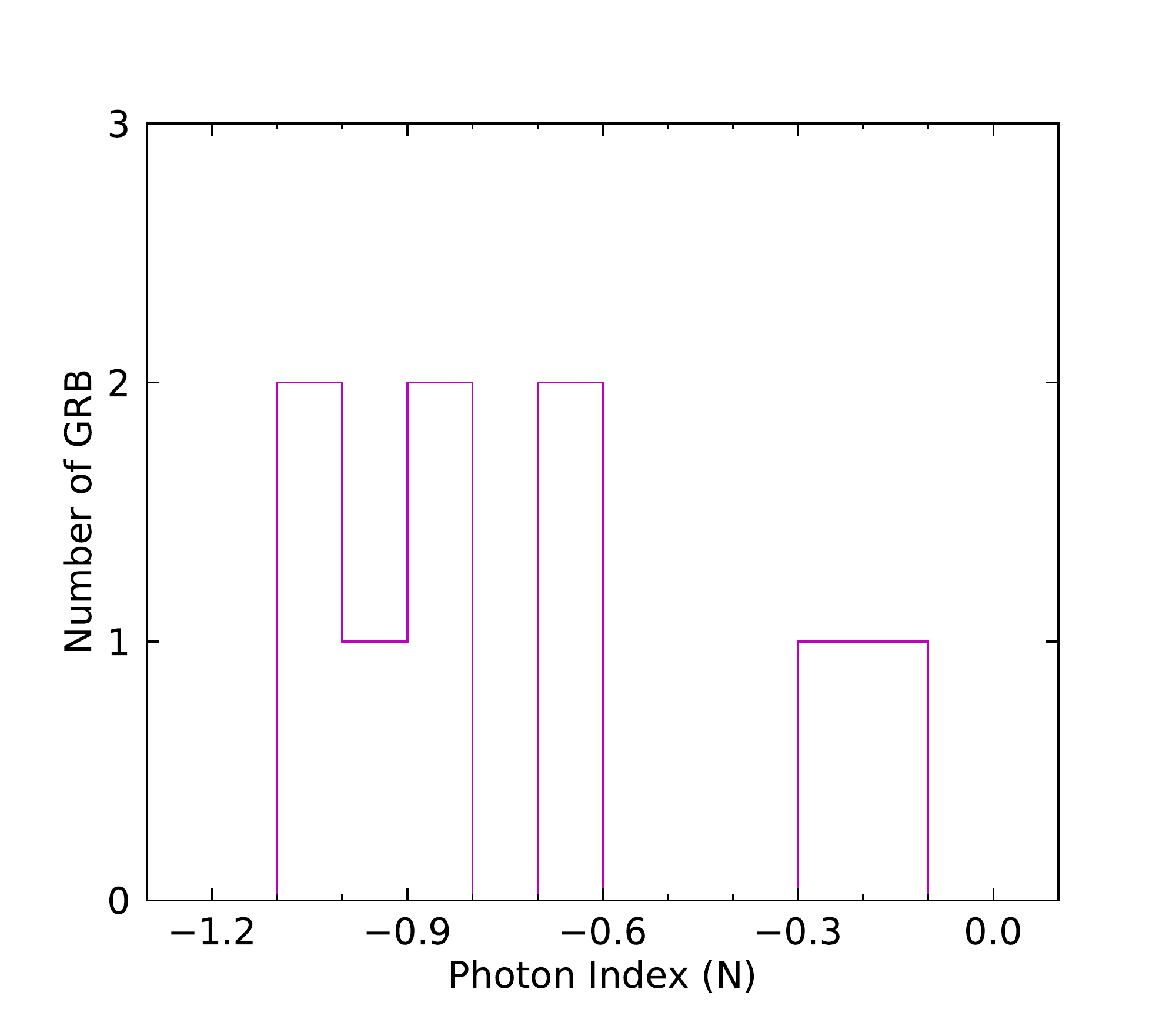}
\includegraphics[scale=0.4]{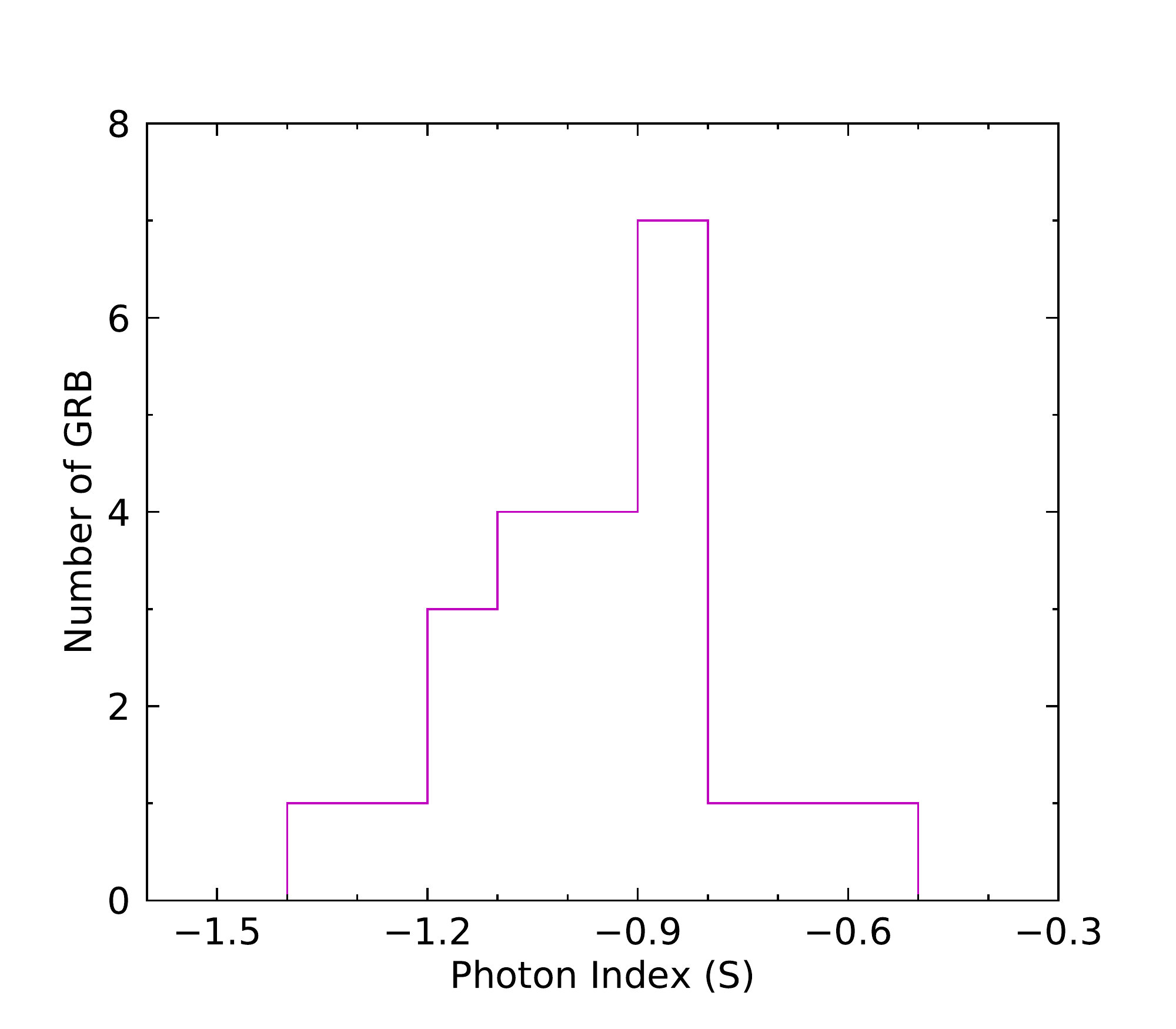}
\includegraphics[scale=0.4]{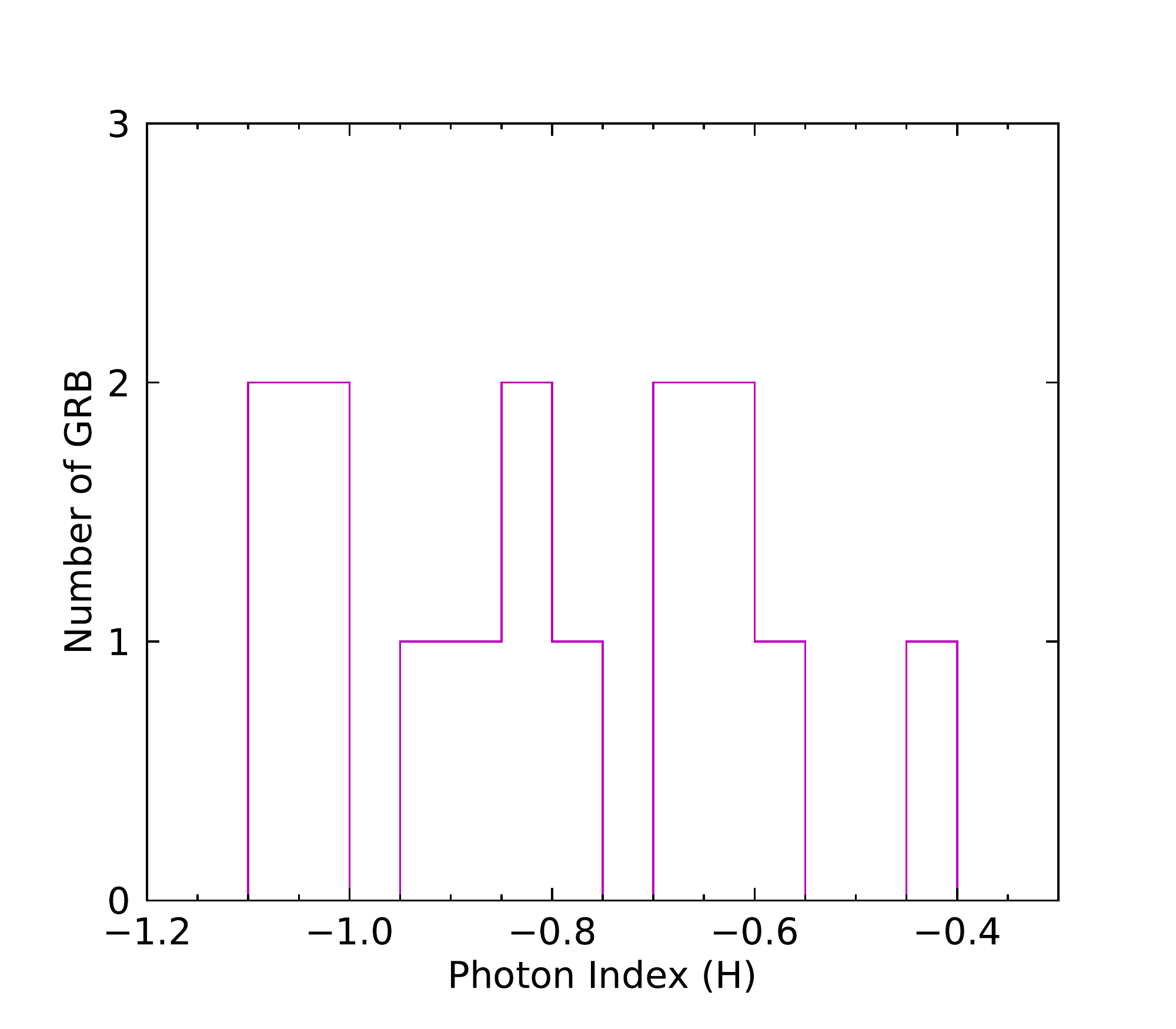}
\caption{GRB low-energy photon index distribution. The GRBs are selected from those with the LAT photon indices obtained from the GBM observing time interval. All notations are the same as those in Figure 5.
\label{fig6}}
\end{figure}

\begin{figure}
\center
\includegraphics[scale=0.4]{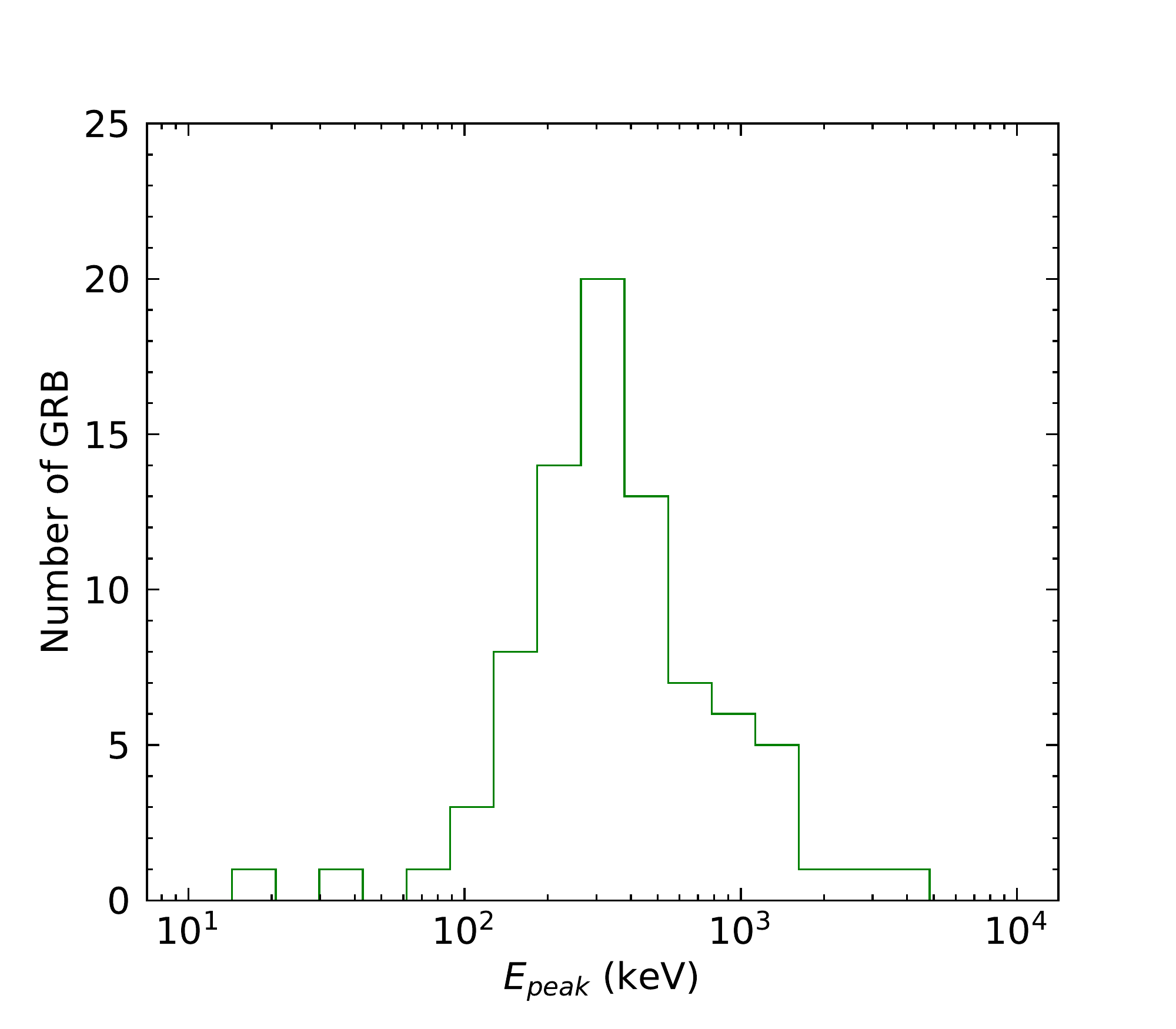}
\includegraphics[scale=0.4]{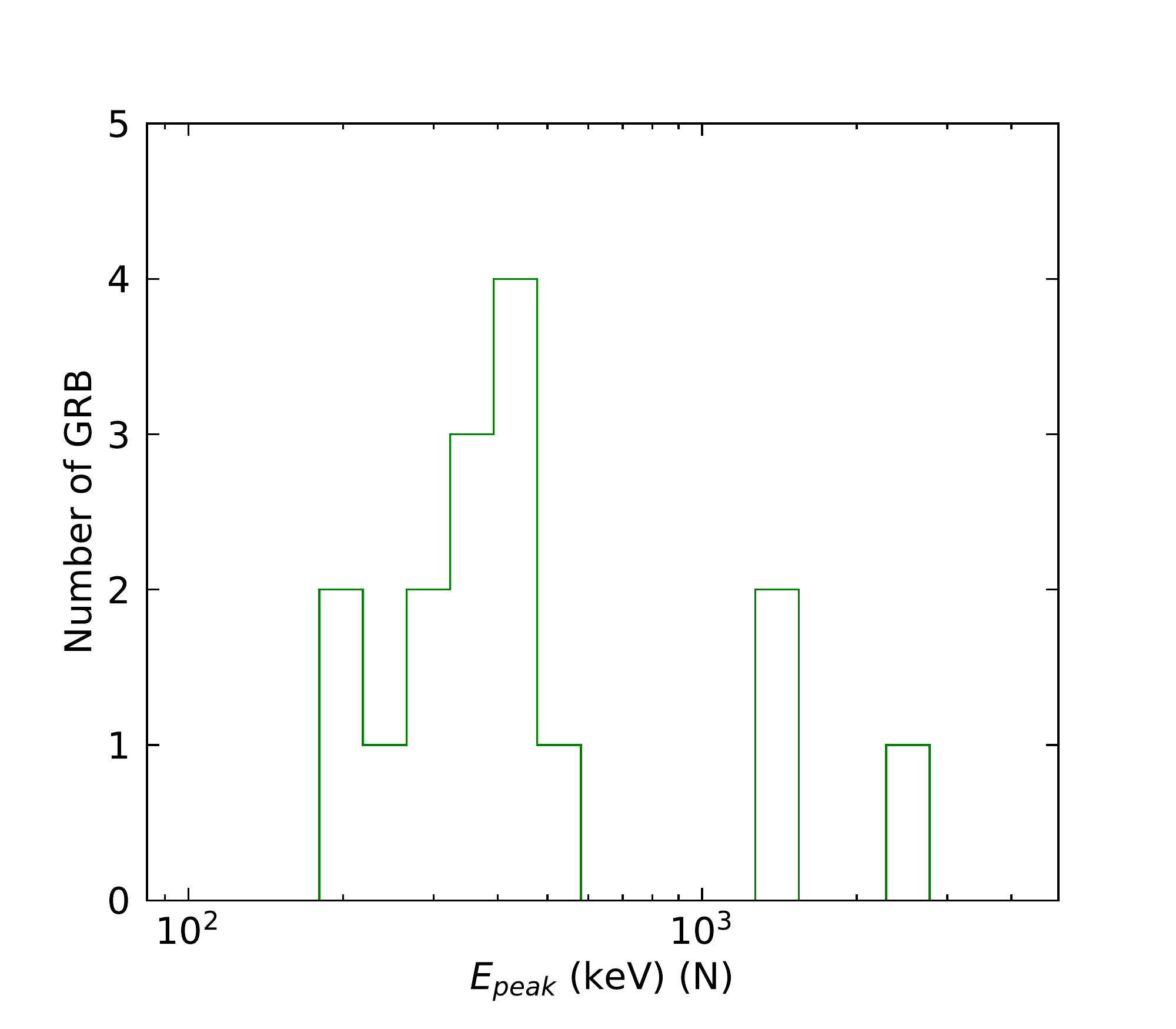}
\includegraphics[scale=0.4]{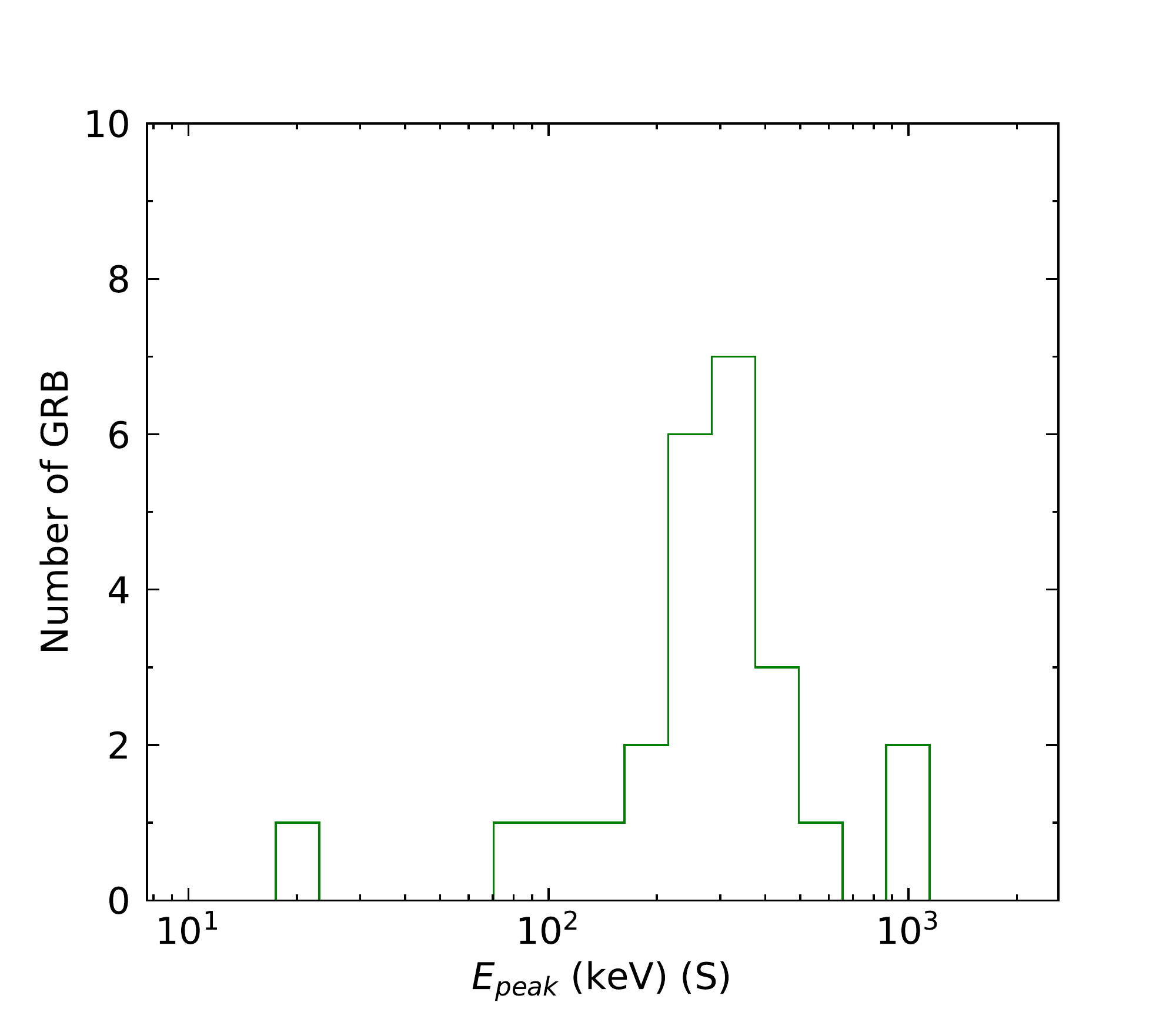}
\includegraphics[scale=0.4]{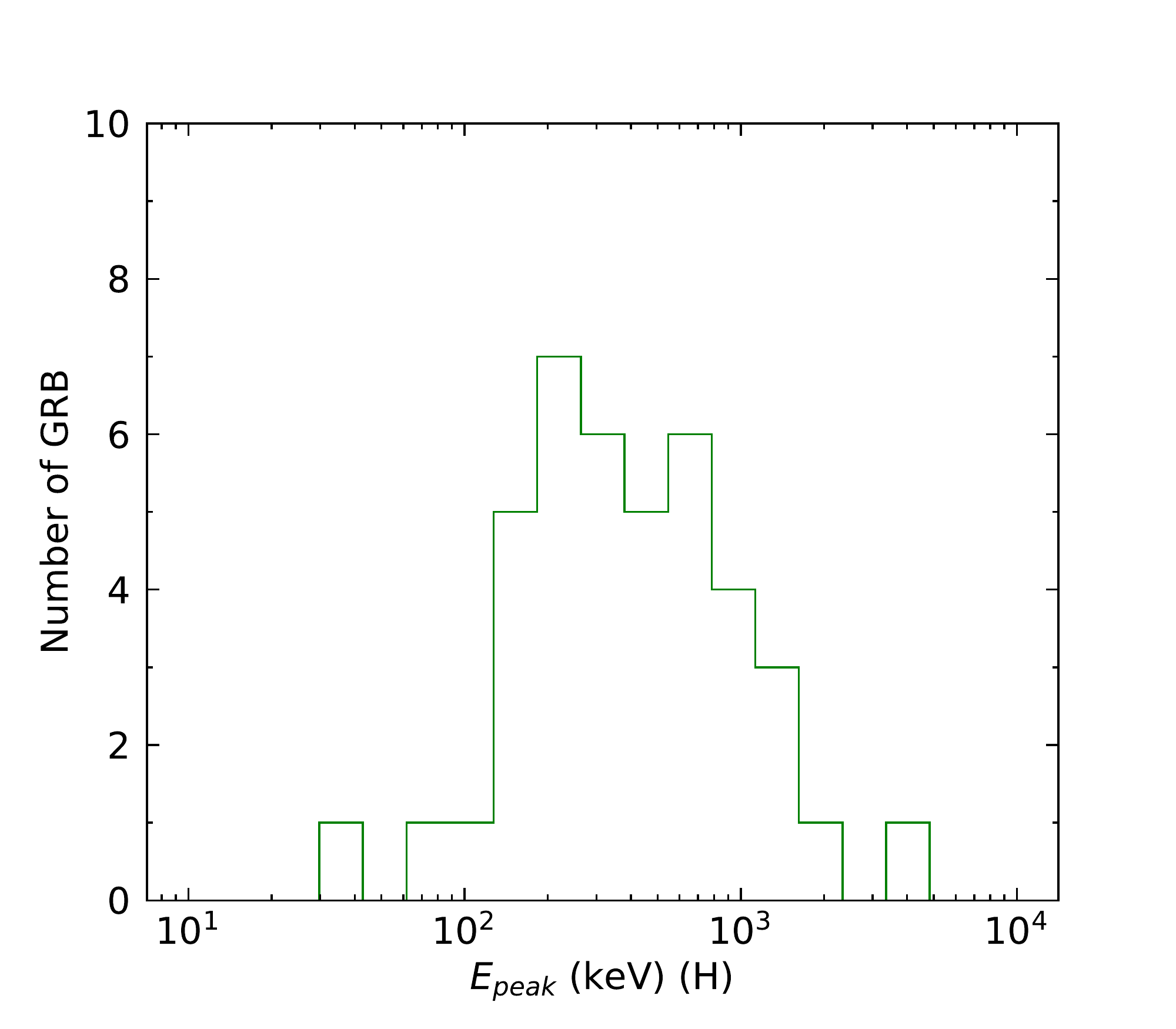}
\caption{$E_{\rm{peak}}$ distribution. The GRBs are selected from those with the LAT photon indices obtained from the LAT observing time interval. Upper left panel: all the GRB $E_{\rm{peak}}$ distributions. Upper right panel: N-GRB $E_{\rm{peak}}$ distribution. Lower left panel: S-GRB $E_{\rm{peak}}$ distribution. Lower right panel: H-GRB $E_{\rm{peak}}$ distribution.
\label{fig7}}
\end{figure}

\begin{figure}
\center
\includegraphics[scale=0.4]{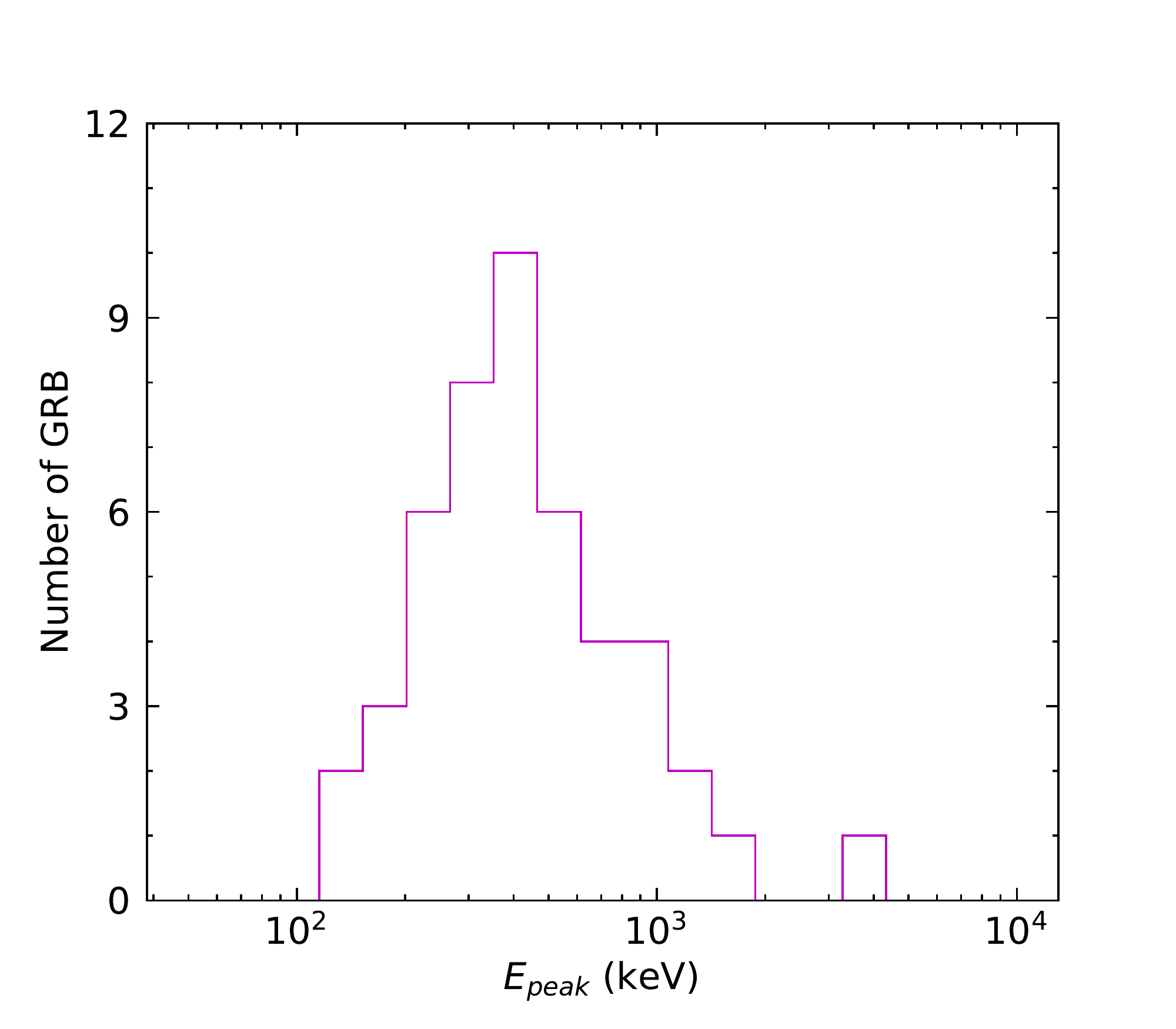}
\includegraphics[scale=0.4]{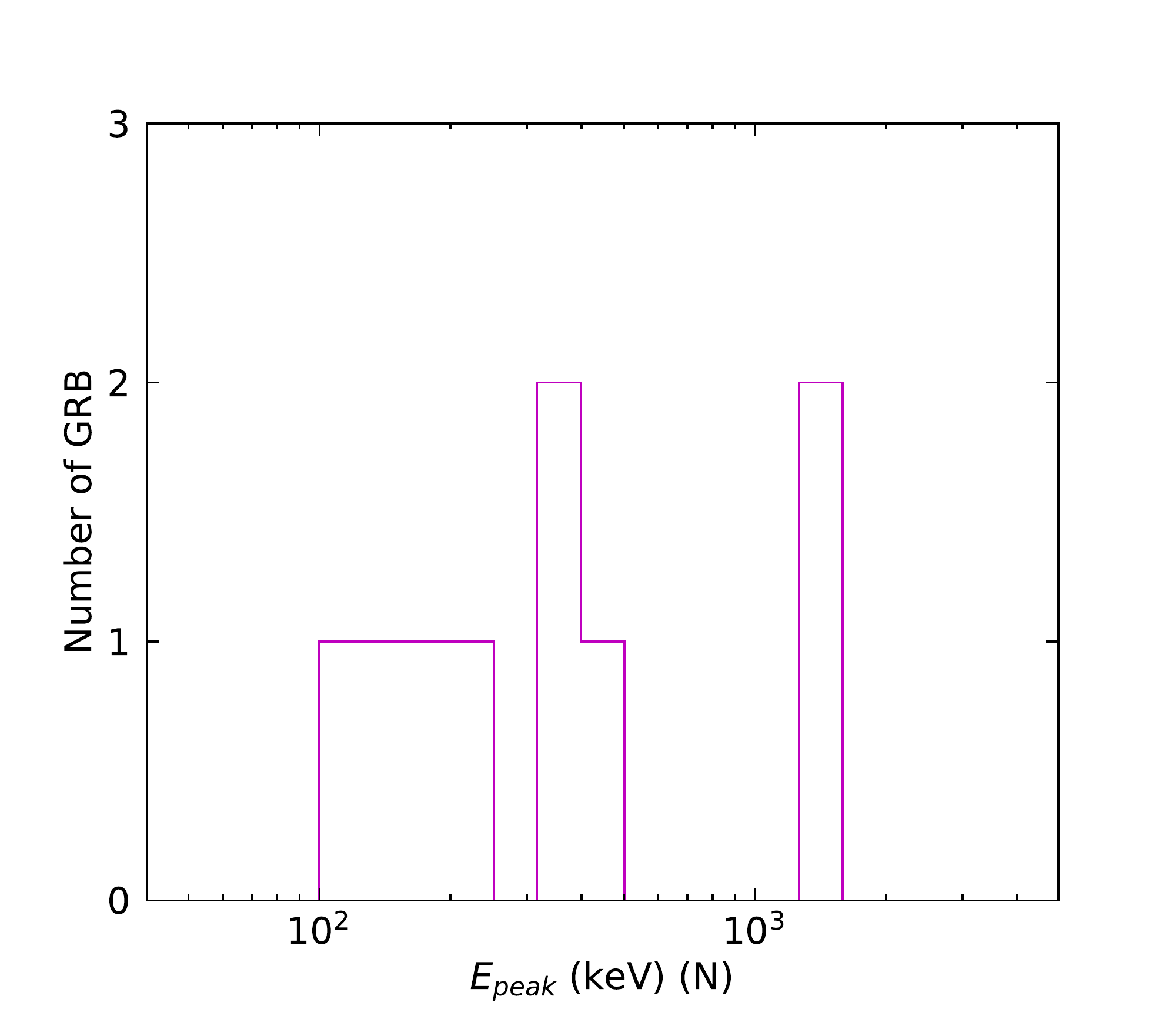}
\includegraphics[scale=0.4]{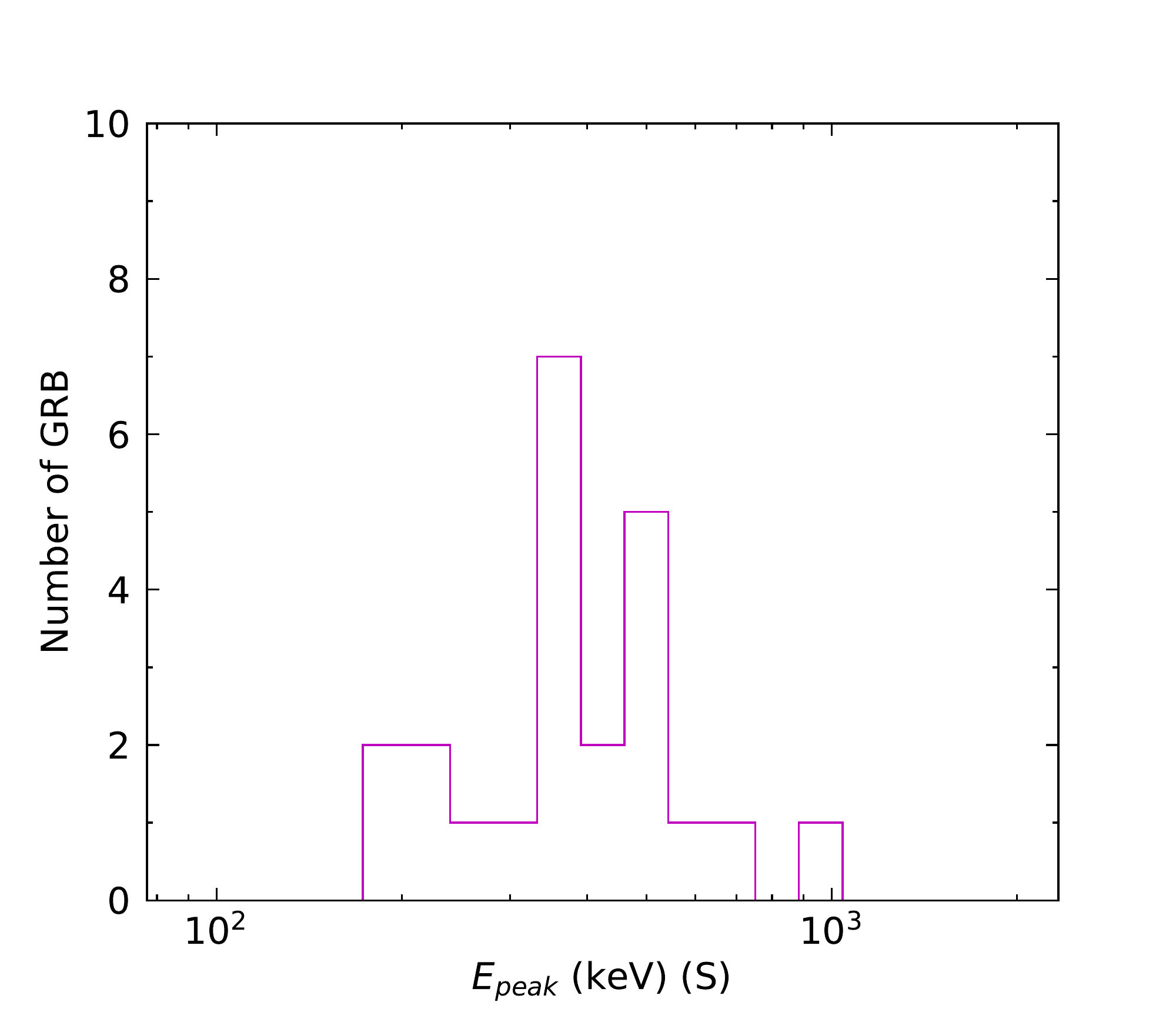}
\includegraphics[scale=0.4]{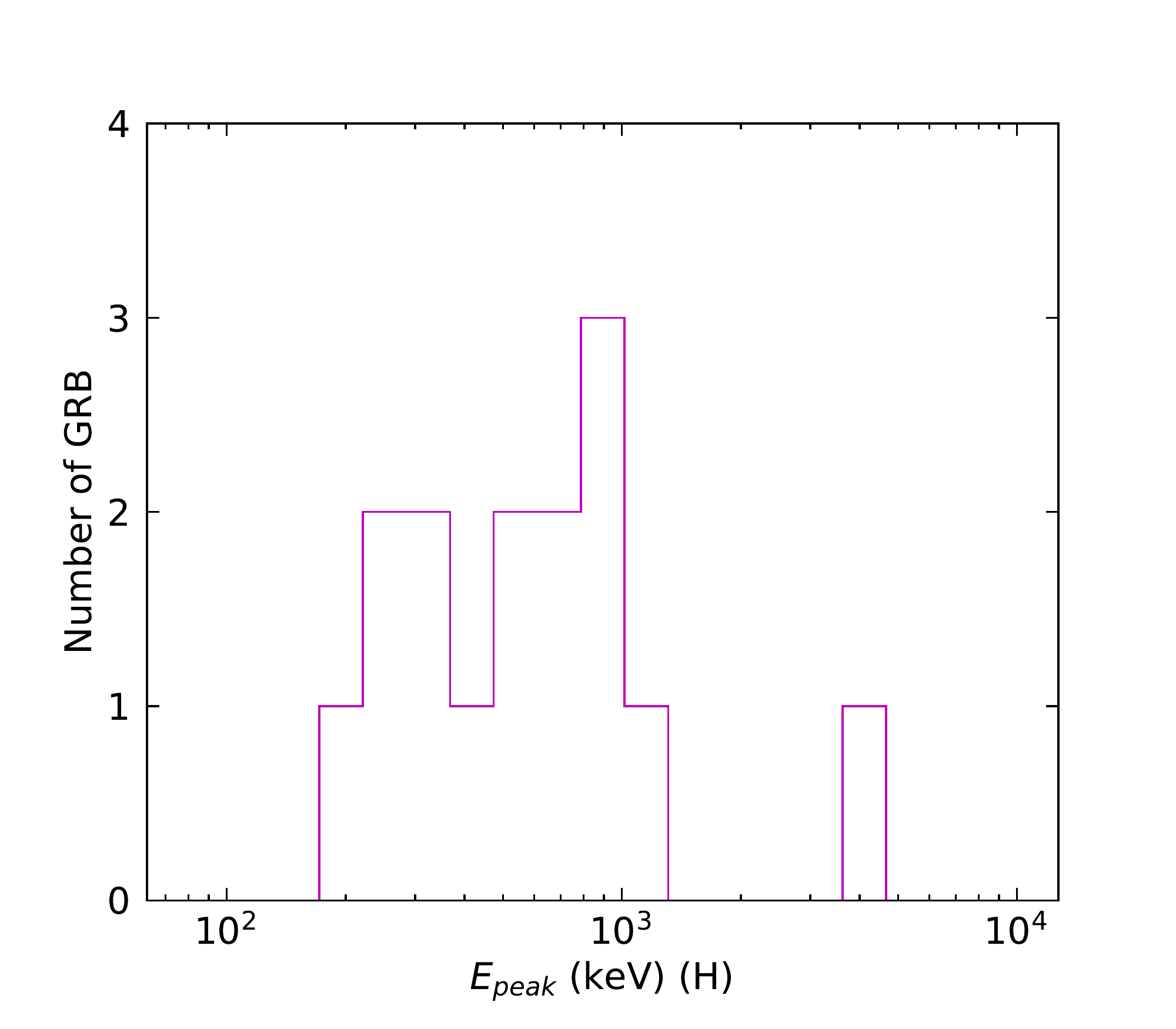}
\caption{$E_{\rm{peak}}$ distribution. The GRBs are selected from those with the LAT photon indices obtained from the GBM observing time interval.
All notations are the same as those in Figure 7.
\label{fig1}}
\end{figure}

\begin{figure}
\center
\includegraphics[scale=0.5]{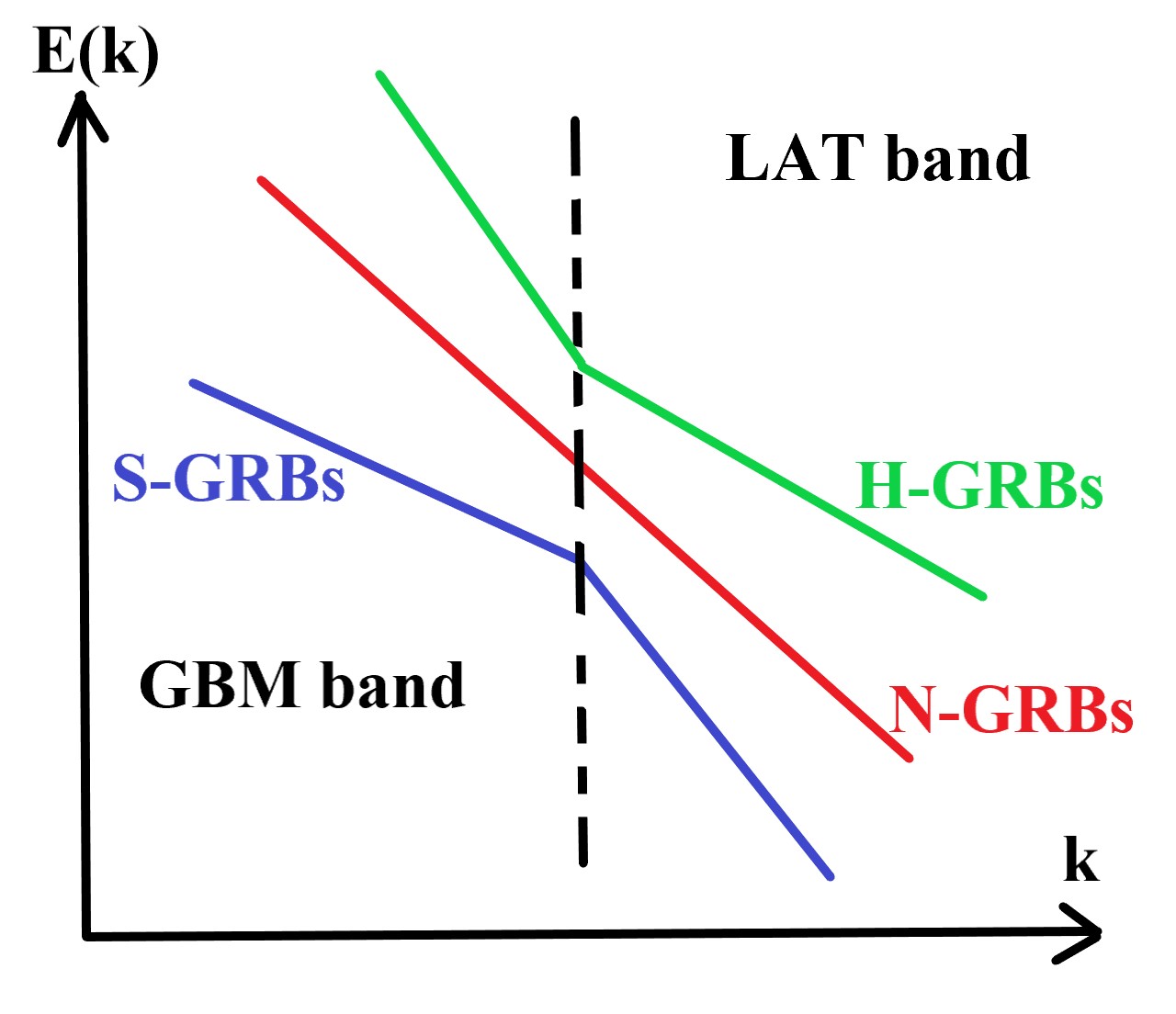}
\caption{Turbulent spectra for three kinds of GRBs. N-GRBs, H-GRBs and S-GRBs have the spectra with the colors of red, green, and blue, respectively. In our model, the radiation in the higher energy band occurs in the smaller lengthscale. GBM and LAT energy bands are separated by the dashed line.
\label{fig3}}
\end{figure}

\begin{deluxetable}{lcc ccc ccccc}
\tablecaption{GRB spectral properties detected by Fermi-LAT, Fermi-GBM, and Swift-BAT \label{tab1}}
\tablewidth{0pt}
\tabletypesize{\scriptsize}
\tablehead{
GRB  & LAT1  & LAT2 & Band $\beta$ & Band $\alpha$ & $E_{\rm{peak}}$($10^{2}$ keV) & BAT PL & BAT CPL
}
\startdata
080825C(S1S2)        & -2.80$\pm$0.40    & -3.80$\pm$0.80    & -2.30$\pm$0.09     & -0.65$\pm$0.03     & 1.74$\pm$0.07       & $-$              & $-$            \\
080916C           & -2.20$\pm$0.06    & -2.22$\pm$0.07    & $-$                & $-$                & $-$                    & $-$              & $-$            \\
081006(N1N2)         & -2.00$\pm$0.30    & -2.20$\pm$0.40    & -2.06$\pm$0.50     & -0.22$\pm$0.45     & 4.62$\pm$1.41       & $-$              & $-$            \\
081009            & -1.80$\pm$0.20    & $-$               & -4.70$\pm$24.71    & -1.58$\pm$0.02     & 0.29$\pm$0.01          & $-$              & $-$            \\
081024B($\ast$)   & -2.20$\pm$0.40    & -2.30$\pm$0.50    & $-$                & -0.94$\pm$0.14     & 17.57$\pm$10.12        & $-$              & $-$            \\
081102B($\ast$)   & -2.50$\pm$0.60    & -2.20$\pm$0.50    & $-$                & -0.81$\pm$0.15     & 6.41$\pm$1.48          & $-$              & $-$            \\
090217            & -2.30$\pm$0.30    & -2.30$\pm$0.30    & $-$                & $-$                & $-$                    & $-$              & $-$            \\
090227A(H1)        & -1.70$\pm$0.40    & $-$               & -1.92$\pm$0.30     & -0.92$\pm$0.06     & 16.91$\pm$4.73        & $-$              & $-$            \\
090228A($\ast$)   & -1.90$\pm$0.50    & $-$               & -3.63$\pm$3.17     & -0.61$\pm$0.03     & 8.51$\pm$0.47          & $-$              & $-$            \\
090323(N1S2)         & -2.30$\pm$0.20    & -3.00$\pm$0.40    & -2.35$\pm$0.15     & -1.18$\pm$0.01     & 4.54$\pm$0.24       & $-$              & $-$            \\
090328(H1S2)         & -2.20$\pm$0.10    & -3.00$\pm$0.50    & -2.39$\pm$0.23     & -1.08$\pm$0.02     & 6.51$\pm$0.44       & $-$              & $-$            \\
090510(H1H2)($\ast$) & -2.05$\pm$0.06    & -1.80$\pm$0.10    & -2.57$\pm$0.37     & -0.84$\pm$0.03     & 42.48$\pm$4.40      & -1.06$\pm$0.27   & -1.06$\pm$0.30 \\
090626(N1)         & -2.10$\pm$0.20    & $-$               & -2.20$\pm$0.08     & -1.24$\pm$0.02     & 1.87$\pm$0.12         & $-$              & $-$            \\
090720B(N1N2)        & -2.30$\pm$0.50    & -2.50$\pm$0.60    & -2.44$\pm$0.42     & -1.04$\pm$0.03     & 12.75$\pm$1.95      & $-$              & $-$            \\
090902B           & -1.94$\pm$0.04    & -1.99$\pm$0.06    & $-$                & -1.01$\pm$0.00     & 10.55$\pm$0.16         & $-$              & $-$            \\
090926A(H1S2)        & -2.14$\pm$0.05    & -2.52$\pm$0.09    & -2.38$\pm$0.05     & -0.85$\pm$0.01     & 3.34$\pm$0.06       & $-$              & $-$            \\
091003(H1H2)         & -1.80$\pm$0.20    & -1.80$\pm$0.40    & -2.21$\pm$0.15     & -1.07$\pm$0.02     & 3.70$\pm$0.27       & $-$              & $-$            \\
091031(N1S2)         & -2.10$\pm$0.20    & -2.70$\pm$0.50    & -2.10$\pm$0.19     & -0.86$\pm$0.04     & 4.56$\pm$0.46       & $-$              & $-$            \\
091120(H1)         & -2.80$\pm$0.40    & $-$               & -3.20$\pm$0.35     & -1.07$\pm$0.02     & 1.25$\pm$0.04         & $-$              & $-$            \\
091127(H1)         & -1.20$\pm$0.40    & $-$               & -2.22$\pm$0.02     & -1.25$\pm$0.07     & 0.35$\pm$0.02         & -2.02$\pm$0.07   & -1.80$\pm$0.28 \\
100116A           & -1.80$\pm$0.20    & -3.20$\pm$0.70    & $-$                & -1.03$\pm$0.02     & 10.74$\pm$0.76         & $-$              & $-$            \\
100213C           & -1.90$\pm$0.40    & $-$               & $-$                & $-$                & $-$                    & $-$              & $-$            \\
100225A           & -2.20$\pm$0.40    & -3.70$\pm$1.00    & $-$                & -0.74$\pm$0.09     & 4.00$\pm$0.61          & $-$              & $-$            \\
100325A           & -1.90$\pm$0.40    & -2.10$\pm$0.40    & $-$     & -0.41$\pm$0.13     & 1.56$\pm$0.13                     & $-$              & $-$            \\
100414A(H1H2)        & -1.80$\pm$0.10    & -2.30$\pm$0.50    & -3.53$\pm$1.25     & -0.62$\pm$0.01     & 6.64$\pm$0.15       & $-$              & $-$            \\
100423B(H1)        & -1.80$\pm$0.60    & $-$               & -2.17$\pm$0.27     & -0.88$\pm$0.05     & 12.00$\pm$1.87        & $-$              & $-$            \\
100511A(S1)        & -1.80$\pm$0.20    & $-$               & -1.60$\pm$0.01     & -0.86$\pm$0.13     & 0.93$\pm$0.18         & $-$              & $-$            \\
100620A           & -4.00$\pm$2.00    & -4.00$\pm$1.00    & $-$                & -1.22$\pm$0.06     & 15.33$\pm$5.99         & $-$              & $-$            \\
100724B(S1S2)        & -4.00$\pm$1.00    & -5.00$\pm$1.00    & -1.97$\pm$0.02     & -0.83$\pm$0.01     & 3.59$\pm$0.10       & $-$              & $-$            \\
100728A(H1)        & -1.60$\pm$0.20    & $-$               & -2.54$\pm$0.10     & -0.51$\pm$0.02     & 2.54$\pm$0.07         & -1.12$\pm$0.03   & -1.06$\pm$0.03 \\
100826A(S1H2)        & -2.30$\pm$0.60    & -1.60$\pm$0.40    & -1.92$\pm$0.02     & -0.81$\pm$0.01     & 2.63$\pm$0.08       & $-$              & $-$            \\
101014A(H1)        & -1.30$\pm$0.30    & $-$               & -2.06$\pm$0.02     & -1.12$\pm$0.01     & 2.02$\pm$0.06         & $-$              & $-$            \\
101107A           & -1.90$\pm$0.30    & -1.90$\pm$0.30    & $-$                & -0.98$\pm$0.09     & 1.43$\pm$0.15          & $-$              & $-$            \\
101227B(H1H2)        & -1.50$\pm$0.50    & -1.60$\pm$0.30    & -2.13$\pm$0.17     & -0.62$\pm$0.17     & 2.13$\pm$0.41       & $-$              & $-$            \\
110120A(H1)        & -2.00$\pm$0.30    & $-$               & -2.83$\pm$1.52     & -0.57$\pm$0.04     & 8.37$\pm$0.72         & $-$              & $-$            \\
110123A(S1)        & -2.20$\pm$0.40    & $-$               & -1.91$\pm$0.05     & -0.53$\pm$0.05     & 2.49$\pm$0.16         & $-$              & $-$            \\
110328B(S1)        & -2.00$\pm$0.30    & $-$               & -1.74$\pm$0.06     & -1.03$\pm$0.06     & 3.61$\pm$0.77         & $-$              & $-$            \\
110428A(H1)        & -1.90$\pm$0.20    & $-$               & -3.09$\pm$0.25     & -0.41$\pm$0.03     & 1.82$\pm$0.05         & $-$              & $-$            \\
110518A           & -1.80$\pm$0.50    & $-$               & $-$                & $-$                & $-$                    & $-$              & $-$            \\
110625A(S1)        & -2.70$\pm$0.30    & $-$               & -2.30$\pm$0.05     & -0.80$\pm$0.02     & 1.65$\pm$0.04         & -1.43$\pm$0.04   & -1.27$\pm$0.13 \\
110721A(S1S2)        & -2.40$\pm$0.20    & -2.60$\pm$0.30    & -1.78$\pm$0.03     & -1.03$\pm$0.02     & 4.65$\pm$0.39       & $-$              & $-$            \\
110728A(S1)($\ast$)& -1.90$\pm$0.40    & $-$               & -1.62$\pm$0.33     & -0.40$\pm$0.49     & 2.66$\pm$0.02         & $-$              & $-$            \\
110731A(N1N2)        & -2.30$\pm$0.20    & -2.60$\pm$0.20    & -2.44$\pm$0.27     & -0.87$\pm$0.03     & 3.22$\pm$0.17       & -1.15$\pm$0.05   & -1.15$\pm$0.05 \\
110903A           & -1.40$\pm$0.30    & -1.50$\pm$0.30    & $-$                & -0.85$\pm$0.03     & 3.08$\pm$0.15          & $-$              & $-$            \\
110921B(H1H2)        & -1.80$\pm$0.30    & -2.00$\pm$0.40    & -2.40$\pm$0.20     & -0.92$\pm$0.02     & 4.95$\pm$0.30       & $-$              & $-$            \\
111210B           & -2.70$\pm$0.50    & $-$               & $-$                & $-$                & $-$                    & $-$              & $-$            \\
120107A           & -1.90$\pm$0.40    & -2.40$\pm$0.50    & $-$                & $-$                & $-$                    & $-$              & $-$            \\
120226A           & -2.90$\pm$0.50    & $-$               & $-$                & $-$                & $-$                    & $-$              & $-$            \\
120316A(H1H2)        & -2.20$\pm$0.30    & -1.90$\pm$0.50    & -2.46$\pm$0.33     & -0.80$\pm$0.03     & 6.52$\pm$0.51       & $-$              & $-$            \\
120420B(S1)        & -2.10$\pm$0.40    & $-$               & -1.64$\pm$0.04     & -0.95$\pm$0.08     & 2.90$\pm$0.68         & $-$              & $-$            \\
120526A(H1)        & -1.80$\pm$0.20    & $-$               & -2.98$\pm$0.24     & -0.83$\pm$0.02     & 7.83$\pm$0.29         & $-$              & $-$            \\
120624B(S1)        & -2.50$\pm$0.10    & $-$               & -2.22$\pm$0.08     & -0.92$\pm$0.01     & 6.35$\pm$0.23         & -1.12$\pm$0.03   & -1.11$\pm$0.04 \\
120709A           & -2.30$\pm$0.20    & -2.20$\pm$0.30    & $-$    & -1.11$\pm$0.03     & 4.91$\pm$0.42                      & $-$              & $-$            \\
120711A(H1)        & -2.10$\pm$0.20    & $-$               & -2.80$\pm$0.09     & -0.98$\pm$0.01     & 13.18$\pm$0.42        & $-$              & $-$            \\
120830A($\ast$)   & -1.90$\pm$0.40    & -2.10$\pm$0.60    & $-$     & -0.35$\pm$0.07     & 10.89$\pm$1.05                    & $-$              & $-$            \\
120911B           & -2.50$\pm$0.20    & -2.50$\pm$0.20    & $-$                & $-$                & $-$                    & $-$              & $-$            \\
121029A           & -1.60$\pm$0.30    & $-$               & $-$                & -0.64$\pm$0.08     & 1.63$\pm$0.09          & $-$              & $-$            \\
121123B           & -2.40$\pm$0.40    & $-$               & $-$                & $-$                & $-$                    & $-$              & $-$            \\
130310A(N1)        & -2.10$\pm$0.30    & $-$               & -2.19$\pm$0.19     & -1.15$\pm$0.02     & 25.01$\pm$4.05        & $-$              & $-$            \\
130325A(H1)        & -1.50$\pm$0.30    & $-$               & -2.05$\pm$0.11     & -0.77$\pm$0.06     & 2.07$\pm$0.22         & $-$              & $-$            \\
130327B           & -1.80$\pm$0.10    & -1.80$\pm$0.30    & $-$                & -0.56$\pm$0.02     & 3.56$\pm$0.08          & $-$              & $-$            \\
130427A(H1H2)        & -1.99$\pm$0.04    & -1.90$\pm$0.05    & -2.83$\pm$0.03     & -1.02$\pm$0.00     & 8.25$\pm$0.05       & -1.18$\pm$0.02   & -1.17$\pm$0.03 \\
130502B(H1H2)        & -2.00$\pm$0.10    & -2.20$\pm$0.20    & -2.51$\pm$0.07     & -0.69$\pm$0.01     & 2.98$\pm$0.05       & $-$              & $-$            \\
130504C(H1S2)        & -1.90$\pm$0.20    & -2.50$\pm$0.50    & -2.12$\pm$0.05     & -1.16$\pm$0.01     & 5.39$\pm$0.23       & $-$              & $-$            \\
130518A(S1S2)        & -2.90$\pm$0.30    & -3.50$\pm$0.70    & -2.18$\pm$0.07     & -0.86$\pm$0.02     & 3.81$\pm$0.15       & $-$              & $-$            \\
130606B(H1)        & -1.70$\pm$0.20    & $-$               & -2.07$\pm$0.02     & -1.11$\pm$0.01     & 4.32$\pm$0.11         & $-$              & $-$            \\
130821A(S1S2)        & -2.40$\pm$0.20    & -2.60$\pm$0.50    & -2.07$\pm$0.04     & -0.89$\pm$0.02     & 2.28$\pm$0.08       & $-$              & $-$            \\
130828A(N1N2)        & -2.20$\pm$0.20    & -2.40$\pm$0.20    & -2.28$\pm$0.09     & -0.14$\pm$0.06     & 1.93$\pm$0.10       & $-$              & $-$            \\
131014A(H1H2)        & -1.90$\pm$0.20    & -2.00$\pm$0.30    & -2.53$\pm$0.03     & -0.43$\pm$0.01     & 3.26$\pm$0.04       & $-$              & $-$            \\
131029A(S1S2)        & -2.40$\pm$0.20    & -2.60$\pm$0.30    & -2.10$\pm$0.16     & -0.94$\pm$0.04     & 2.24$\pm$0.18       & $-$              & $-$            \\
131108A(S1S2)        & -2.70$\pm$0.10    & -2.60$\pm$0.10    & -2.46$\pm$0.19     & -0.91$\pm$0.02     & 3.67$\pm$0.16       & $-$              & $-$            \\
131209A           & -3.30$\pm$0.70    & $-$               & -4.03$\pm$81.50    & -0.59$\pm$0.05     & 2.99$\pm$0.15          & $-$              & $-$            \\
131231A(H1S2)        & -1.70$\pm$0.10    & -2.70$\pm$0.70    & -2.30$\pm$0.03     & -1.22$\pm$0.01     & 1.78$\pm$0.04       & $-$              & $-$            \\
140102A(H1)        & -2.10$\pm$0.30    & $-$               & -2.71$\pm$0.17     & -0.87$\pm$0.02     & 1.91$\pm$0.06         & -1.37$\pm$0.04   & -1.23$\pm$0.15 \\
140104B(S1)        & -2.00$\pm$0.20    & $-$               & -1.58$\pm$0.01     & -1.37$\pm$34.70    & 0.19$\pm$75.90        & $-$              & $-$            \\
140110A(N1N2)        & -2.60$\pm$0.30    & -2.60$\pm$0.30    & -2.46$\pm$0.69     & -0.66$\pm$0.07     & 14.62$\pm$3.07      & $-$              & $-$            \\
140206B(N1S2)        & -2.10$\pm$0.10    & -2.80$\pm$0.30    & -2.04$\pm$0.05     & -1.40$\pm$0.01     & 4.76$\pm$0.27       & $-$              & $-$            \\
140402A($\ast$)   & -1.80$\pm$0.30    & -2.40$\pm$0.60    & $-$    & -0.14$\pm$0.29     & 10.93$\pm$2.33                     & -1.15$\pm$0.58   & -1.17$\pm$0.72 \\
140523A(H1H2)        & -2.00$\pm$0.10    & -1.90$\pm$0.20    & -2.70$\pm$0.21     & -1.07$\pm$0.01     & 2.64$\pm$0.09       & $-$              & $-$            \\
140528A           & -2.00$\pm$0.30    & $-$               & -3.40$\pm$9.08     & -0.73$\pm$0.03     & 2.62$\pm$0.10          & $-$              & $-$            \\
140619B           & -1.90$\pm$0.20    & -2.00$\pm$0.20    & $-$    & -0.18$\pm$0.22     & 13.07$\pm$2.97                     & $-$              & $-$            \\
140723A(H1H2)        & -2.20$\pm$0.20    & -2.20$\pm$0.20    & -2.47$\pm$1.28     & -1.05$\pm$0.05     & 11.37$\pm$2.12      & $-$              & $-$            \\
140729A           & -1.80$\pm$0.30    & -1.90$\pm$0.30    & $-$                & -0.42$\pm$0.09     & 7.87$\pm$0.92          & $-$              & $-$            \\
140810A           & -1.50$\pm$0.20    & $-$               & $-$                & $-$                & $-$                    & $-$              & $-$            \\
140928A(H1)        & -1.90$\pm$0.50    & $-$               & -2.66$\pm$0.60     & 0.08$\pm$0.14      & 6.93$\pm$0.78         & $-$              & $-$            \\
141012A           & -2.00$\pm$0.20    & -2.40$\pm$0.40    & $-$                & $-$                & $-$                    & $-$              & $-$            \\
141028A(S1S2)        & -2.40$\pm$0.20    & -3.20$\pm$0.50    & -1.97$\pm$0.05     & -0.84$\pm$0.03     & 2.93$\pm$0.18       & $-$              & $-$            \\
141102A           & -1.90$\pm$0.40    & $-$               & $-$                & $-$                & $-$                    & $-$              & $-$            \\
141113A($\ast$)   & -2.60$\pm$0.70    & -2.10$\pm$0.50    & $-$                & $-$                & $-$                    & $-$              & $-$            \\
141207A(H1H2)        & -1.80$\pm$0.10    & -1.90$\pm$0.10    & -2.86$\pm$0.39     & -0.70$\pm$0.03     & 9.85$\pm$0.54       & $-$              & $-$            \\
141221B           & -1.60$\pm$0.40    & $-$               & $-$                & $-$                & $-$                    & $-$              & $-$            \\
141222A           & -2.10$\pm$0.30    & $-$               & $-$                & $-$                & $-$                    & $-$              & $-$            \\
150210A           & -2.20$\pm$0.30    & -2.20$\pm$0.30    & $-$                & -1.07$\pm$0.01     & 29.45$\pm$2.21         & $-$              & $-$            \\
150314A(N1)        & -2.50$\pm$0.40    & $-$               & -2.60$\pm$0.10     & -0.68$\pm$0.01     & 3.47$\pm$0.08         & -1.08$\pm$0.03   & -1.00$\pm$0.07 \\
150403A(H1)        & -1.70$\pm$0.30    & $-$               & -2.11$\pm$0.06     & -0.87$\pm$0.02     & 4.29$\pm$0.21         & -1.22$\pm$0.04   & -1.10$\pm$0.15 \\
150510A           & -2.00$\pm$0.30    & -2.40$\pm$0.50    & $-$                & $-$                & $-$                    & $-$              & $-$            \\
150514A(H1)        & -1.00$\pm$0.40    & $-$               & -2.43$\pm$0.18     & -1.21$\pm$0.10     & 0.65$\pm$0.06         & $-$              & $-$            \\
150523A(H1H2)        & -1.90$\pm$0.10    & -2.10$\pm$0.20    & -3.27$\pm$3.89     & -0.56$\pm$0.03     & 5.45$\pm$0.21       & $-$              & $-$            \\
150627A(H1)        & -1.70$\pm$0.10    & $-$               & -2.18$\pm$0.03     & -1.01$\pm$0.01     & 2.25$\pm$0.05         & $-$              & $-$            \\
150702A           & -2.20$\pm$0.40    & $-$               & $-$                & $-$                & $-$                    & $-$              & $-$            \\
150902A(N1S2)        & -2.30$\pm$0.20    & -3.10$\pm$0.40    & -2.25$\pm$0.04     & -0.54$\pm$0.02     & 3.46$\pm$0.08       & $-$              & $-$            \\
160325A(N1)        & -2.40$\pm$0.20    & $-$               & -2.32$\pm$0.15     & -0.75$\pm$0.04     & 2.40$\pm$0.14         & -1.27$\pm$0.04   & -1.12$\pm$0.12 \\
160503A           & -3.10$\pm$0.50    & $-$               & $-$                & $-$                & $-$                    & $-$              & $-$            \\
160509A(S1S2)        & -2.40$\pm$0.10    & -2.40$\pm$0.10    & -2.23$\pm$0.05     & -1.02$\pm$0.01     & 3.55$\pm$0.10       & $-$              & $-$            \\
160521B(H1)        & -1.40$\pm$0.30    & $-$               & -2.46$\pm$0.09     & -0.50$\pm$0.04     & 1.49$\pm$0.05         & $-$              & $-$            \\
160623A(S1)        & -2.00$\pm$0.10    & $-$               & -1.61$\pm$0.32     & -1.49$\pm$0.17     & 9.98$\pm$46.25        & $-$              & $-$            \\
160625B(S1S2)        & -2.35$\pm$0.07    & -2.33$\pm$0.08    & -2.18$\pm$0.02     & -0.93$\pm$0.00     & 4.71$\pm$0.06       & $-$              & $-$            \\
160709A           & -2.40$\pm$0.30    & -2.40$\pm$0.30    & $-$                & $-$                & $-$                    & -1.11$\pm$0.28   & -0.66$\pm$0.27 \\
160816A(H1N2)        & -2.20$\pm$0.20    & -2.70$\pm$0.40    & -2.89$\pm$0.21     & -0.70$\pm$0.02     & 2.28$\pm$0.05       & $-$              & $-$            \\
160821A(S1S2)        & -2.60$\pm$0.20    & -5.10$\pm$0.70    & -2.31$\pm$0.03     & -1.05$\pm$0.00     & 9.41$\pm$0.16       & -1.53$\pm$0.06   & -1.47$\pm$0.15 \\
160829A($\ast$)   & -1.30$\pm$0.30    & $-$               & $-$                & $-$                & $-$                    & $-$              & $-$            \\
160905A(H1H2)        & -1.80$\pm$0.20    & -1.70$\pm$0.30    & -2.96$\pm$0.29     & -0.87$\pm$0.01     & 10.03$\pm$0.41      & -0.96$\pm$0.03   & -0.96$\pm$0.03 \\
161015A(S1N2)        & -2.70$\pm$0.60    & -2.20$\pm$0.40    & -2.39$\pm$0.15     & -0.85$\pm$0.04     & 1.51$\pm$0.08       & $-$              & $-$            \\
161109A           & -2.20$\pm$0.30    & $-$               & $-$                & $-$                & $-$                    & $-$              & $-$            \\
170115B           & -2.70$\pm$0.40    & -3.20$\pm$0.50    & $-$                & $-$                & $-$                    & $-$              & $-$            \\
170127C($\ast$)   & -2.90$\pm$0.50    & $-$               & $-$                & $-$                & $-$                    & $-$              & $-$            \\
170214A(N1S2)        & -2.45$\pm$0.09    & -2.70$\pm$0.10    & -2.51$\pm$0.10     & -0.98$\pm$0.01     & 4.81$\pm$0.11       & $-$              & $-$            \\
170306B           & -2.80$\pm$0.90    & $-$               & $-$                & $-$                & $-$                    & $-$              & $-$            \\
170329A           & -2.20$\pm$0.30    & -2.10$\pm$0.40    & $-$                & $-$                & $-$                    & $-$              & $-$            \\
170405A(N1S2)        & -2.80$\pm$0.30    & -5.00$\pm$2.00    & -2.35$\pm$0.09     & -0.80$\pm$0.02     & 2.67$\pm$0.09       & -1.51$\pm$0.12   & -1.52$\pm$0.14 \\
170409A           & -2.10$\pm$0.30    & $-$               & $-$                & $-$                & $-$                    & $-$              & $-$            \\
170424A           & -2.00$\pm$0.40    & -1.90$\pm$0.50    & $-$                & $-$                & $-$                    & $-$              & $-$            \\
170510A(N1N2)        & -2.30$\pm$0.40    & -2.20$\pm$0.40    & -2.28$\pm$0.16     & -1.03$\pm$0.02     & 3.66$\pm$0.24       & $-$              & $-$            \\
170522A           & -1.60$\pm$0.20    & -1.80$\pm$0.30    & $-$                & $-$                & $-$                    & $-$              & $-$            \\
170808B           & -2.20$\pm$0.30    & $-$               & $-$                & $-$                & $-$                    & $-$              & $-$            \\
170906A(N1)        & -2.10$\pm$0.10    & $-$               & -2.20$\pm$0.07     & -1.03$\pm$0.02     & 2.74$\pm$0.11         & -1.34$\pm$0.02   & -1.29$\pm$0.07 \\
171010A(H1)        & -2.00$\pm$0.10    & $-$               & -2.19$\pm$0.01     & -1.09$\pm$0.01     & 1.38$\pm$0.01         & $-$              & $-$            \\
171102A           & -2.60$\pm$0.60    & $-$               & $-$                & $-$                & $-$                    & $-$              & $-$            \\
171120A(S1N2)        & -2.20$\pm$0.20    & -2.10$\pm$0.40    & -2.06$\pm$0.06     & -0.91$\pm$0.06     & 1.21$\pm$0.10       & -1.53$\pm$0.06   & -1.08$\pm$0.27 \\
171124A           & -2.20$\pm$0.20    & -2.00$\pm$0.20    & $-$    & -0.94$\pm$0.04     & 10.84$\pm$1.48                     & $-$              & $-$            \\
171210A           & -2.40$\pm$0.30    & $-$               & $-$                & $-$                & $-$                    & $-$              & $-$            \\
180210A(H1)        & -1.80$\pm$0.10    & $-$               & -2.56$\pm$0.06     & -0.57$\pm$0.02     & 1.42$\pm$0.03         & $-$              & $-$            \\
180305A(H1)        & -2.00$\pm$1.00    & $-$               & -2.24$\pm$0.08     & -0.41$\pm$0.03     & 3.13$\pm$0.11         & $-$              & $-$            \\
180526A           & -1.80$\pm$0.30    & $-$               & $-$                & $-$                & $-$                    & $-$              & $-$            \\
180703A(S1S2)        & -2.50$\pm$0.30    & -3.00$\pm$1.00    & -1.97$\pm$0.10     & -0.78$\pm$0.04     & 3.51$\pm$0.32       & $-$              & $-$            \\
180703B($\ast$)   & -1.70$\pm$0.40    & $-$               & $-$                & $-$                & $-$                    & $-$              & $-$            \\
180718B           & -2.90$\pm$0.60    & -2.80$\pm$0.50    & $-$                & $-$                & $-$                    & $-$              & $-$            \\
180720B(H1S2)        & -2.23$\pm$0.10    & -3.20$\pm$0.40    & -2.49$\pm$0.07     & -1.17$\pm$0.00     & 6.36$\pm$0.15       & -1.35$\pm$0.03   & -1.34$\pm$0.04 \\
\enddata
\tablecomments{``LAT1" and ``LAT2" indicate the LAT photon indices obtained from the LAT and GBM observing time intervals, respectively. Band function is used to determine the $E_{\rm{peak}}$, the low-energy photon index (indicated by ``Band $\alpha$"), and the high-energy photon index (indicated by ``Band $\beta$") for GBM spectrum. ``BAT PL" and ``BAT CPL" indicate the BAT photon index obtained by the power-law and the cutoff power-law fittings, respectively. In this paper, we take the ``Band $\beta$¡± to be the GBM photon index and compare to the LAT photon index. GRB names with the capital letters of N, S, and H indicate N-GRBs, S-GRBs, and H-GRBs, respectively. ``1" and ``2" behind the letters indicate that the classification is based on the LAT photon indices obtained from the LAT and GBM observing time intervals, respectively. Short GRBs are labeled by the symbol $\ast$.}
\end{deluxetable}






\end{document}